%% file: template.tex
\documentclass[journal]{vgtc}                     % final (journal style)}
\DeclareUnicodeCharacter{200B}{}
\usepackage{subcaption}
\usepackage[percent]{overpic} % in preamble
\usepackage{pifont}

\onlineid{1444}

\vgtccategory{IEEE VIS Full Paper}

\vgtcpapertype{Representations \& Interaction}

\title{NLI4VolVis: Natural Language Interaction for Volume Visualization via LLM Multi-Agents and Editable 3D Gaussian Splatting}

\author{%
  \authororcid{Kuangshi Ai}{0009-0005-7171-6529}, 
  \authororcid{Kaiyuan Tang}{0009-0001-3512-0112}, and
  \authororcid{Chaoli Wang}{0000-0002-0859-3619}
}

\authorfooter{
  \item The authors are with %the University of Notre Dame. 
the Department of Computer Science and Engineering, University of Notre Dame, Notre Dame, IN 46556, USA.\\
E-mail: \{kai, ktang2, chaoli.wang\}@nd.edu.
}

\abstract{Traditional volume visualization (VolVis) methods, like direct volume rendering, suffer from rigid transfer function designs and high computational costs. Although novel view synthesis approaches enhance rendering efficiency, they require additional learning effort for non-experts and lack support for semantic-level interaction. To bridge this gap, we propose NLI4VolVis, an interactive system that enables users to explore, query, and edit volumetric scenes using natural language. NLI4VolVis integrates multi-view semantic segmentation and vision-language models to extract and understand semantic components in a scene. We introduce a multi-agent large language model architecture equipped with extensive function-calling tools to interpret user intents and execute visualization tasks. The agents leverage external tools and declarative VolVis commands to interact with the VolVis engine powered by 3D editable Gaussians, enabling open-vocabulary object querying, real-time scene editing, best-view selection, and 2D stylization. We validate our system through case studies and a user study, highlighting its improved accessibility and usability in volumetric data exploration. We strongly recommend readers check our case studies, demo video, and source code at \url{https://nli4volvis.github.io/}.}


\teaser{
	\hspace*{-0.05\linewidth}
	\includegraphics[width=1.1\linewidth]{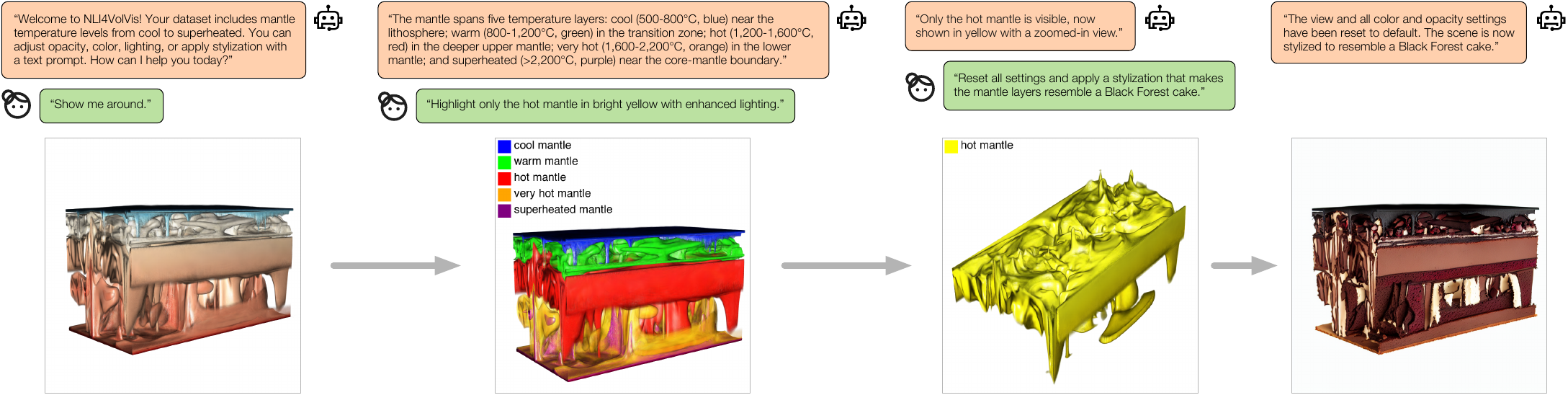}
	\vspace{-0.25in}
	\caption{An example of chatting with NLI4VolVis about the mantle temperature dataset. Our system lets users intuitively explore and edit volumetric datasets using natural language. Through open-vocabulary querying, real-time scene editing, and stylization, users can easily highlight semantic structures, adjust visualization properties, and apply creative styles.} 
	\label{fig:teaser}
}

\graphicspath{{figs/}{figures/}{pictures/}{images/}{./}} % where to search for the images

\usepackage{tabu}                      % only used for the table example
\usepackage{booktabs}                  % only used for the table example
\usepackage{lipsum}                    % used to generate placeholder text
\usepackage{mwe}                       % used to generate placeholder figures

\usepackage{mathptmx}                  % use matching math font
\usepackage{multirow}
\usepackage{graphicx}
\usepackage{times}
\usepackage{algorithm}
\usepackage{algorithmic}
\usepackage{amsfonts}
\usepackage{booktabs}         % only used for the table example
\usepackage{gensymb}
\usepackage{amsmath}
\usepackage{comment}
\usepackage{times} % we use Times as the main font
\usepackage{caption}
\usepackage{bm}
\usepackage{multirow}
\usepackage{color} %ref: http://en.wikibooks.org/wiki/LaTeX/Colors
\usepackage{anyfontsize}
\usepackage{soul}

\newcommand{\hot}[1]{{\color{black} #1}}

\newenvironment{myitemize}{
\begin{itemize}
 \setlength{\itemsep}{1pt}
 \setlength{\parskip}{0pt}
 \setlength{\parsep}{0pt}}{\end{itemize}
 
}

\begin{document}

%%%%%%%%%%%%%%%%%%%%%%%%%%%%%%%%%%%%%%%%%%%%%%%%%%%%%%%%%%%%%%%%
%%%%%%%%%%%%%%%%%%%%%% START OF THE PAPER %%%%%%%%%%%%%%%%%%%%%%
%%%%%%%%%%%%%%%%%%%%%%%%%%%%%%%%%%%%%%%%%%%%%%%%%%%%%%%%%%%%%%%%

\firstsection{Introduction}

\maketitle
\input{intro}

\vspace{-0.05in}
\input{related}

\vspace{-0.05in}
\input{method}

\vspace{-0.05in}
\input{results}

\vspace{-0.05in}
\input{conclusion}

% if specified like this the section will be ommitted in review mode
\vspace{-0.05in}
\acknowledgments{This research was supported in part by the U.S.\ National Science Foundation through grants IIS-1955395, IIS-2101696, OAC-2104158, and IIS-2401144, and the U.S.\ Department of Energy through grant DE-SC0023145. The authors would like to thank the anonymous reviewers for their insightful comments.}

\vspace{-0.05in}
\input{appendix}

\vspace{-0.05in}
\bibliographystyle{abbrv-doi-hyperref}

\bibliography{template}

\end{document}

%% file: intro.tex
\begin{comment}

{\em volume visualization} (VolVis)\\
{\em direct volume rendering} (DVR)\\
{\em transfer function} (TF)\\
{\em novel view synthesis} (NVS)\\
{\em 3D Gaussian splatting} (3DGS)\\
%{\em open-vocabulary segmentation} (OVS)\\
{\em natural language interaction} (NLI)\\
{\em large language model} (LLM)

\end{comment}

\hot{
In {\em scientific visualization} (SciVis), {\em volume visualization} (VolVis) enables domain experts to explore and analyze complex 3D volumetric datasets. 
Traditionally, {\em direct volume rendering} (DVR) has been the dominant approach, valued for its ability to produce high-quality visualizations via ray marching through dense voxel grids. 
However, DVR suffers from significant limitations: it is computationally intensive, requires substantial storage, and scales poorly with resolution, often rendering real-time interaction impractical.

To overcome these challenges, researchers have turned to {\em novel view synthesis} (NVS) techniques that generate new views from sparsely sampled rendering images, eliminating the need for full volumetric integration. 
Among these techniques, {\em 3D Gaussian splatting} (3DGS)~\cite{Fei-TVCG24} has emerged as a promising alternative. 
3DGS directly rasterizes point-based Gaussian primitives onto the screen, offering rapid rendering and compact scene representations well suited to modern GPUs. 
This approach enables more efficient and interactive visualization pipelines and serves as a foundation for flexible volumetric scene editing.

Despite such rendering improvements, the creation of meaningful visualizations still hinges on the {\em transfer function} (TF), which maps voxel intensities to color and opacity. 
While many advanced TF techniques have been proposed~\cite{Ljung-CGF16, Kniss-VIS01, Kindlmann-VIS03, Bruckner-CGF07, Caban-TVCG08, Correa-TVCG08} to enhance expressiveness, TF design remains a complex and often unintuitive task. 
Conventional TFs, in particular, lack semantic awareness, making object-level exploration and manipulation challenging in intricate datasets.

Recent advances in vision-language models have introduced automatic, text-driven TF generation~\cite{Jeong-VISSP24}. 
However, these methods still fall short of enabling interactive, real-time scene editing through natural language, underscoring a persistent disconnect between {\em natural language interaction} (NLI) and intuitive, semantic-level control in VolVis.

The rise of {\em large language models} (LLMs) presents a transformative opportunity to bridge this gap. 
With strong reasoning and planning capabilities, LLMs can act as intelligent agents—interpreting user intent, orchestrating multi-step tasks, and generating context-aware responses. 
When combined with vision-language models, they facilitate semantic comprehension of visual data, enabling open-vocabulary interaction and editing. 
Moreover, {\em multi-agent systems}—comprising collaborative LLMs—can support distributed task execution, enhancing system robustness and flexibility.

Yet, no existing system has unified editable volumetric representations, open-vocabulary scene understanding, and collaborative multi-agent LLMs to support intuitive, natural language-based VolVis. 
Prior efforts~\cite{Jeong-VISSP24, Jia-arXiv23} either lack object-level semantic granularity or fall short of enabling real-time interaction and flexible scene manipulation.

To address these gaps, we present NLI4VolVis (\underline{N}atural \underline{L}anguage \underline{I}nteraction for \underline{Vol}ume \underline{Vis}ualization), a unified system that seamlessly integrates rendering, perception, and interaction. 
NLI4VolVis combines:
(1) editable 3DGS~\cite{Tang-PVIS25} for real-time volumetric rendering,
(2) vision-language models for semantic scene understanding, and
(3) a multi-agent LLM-based framework to interpret user commands and manipulate volumetric scenes interactively.
With NLI4VolVis, users across all levels of expertise can explore, query, and edit scenes using natural language. 
The system supports semantic-level operations such as adjusting color and opacity, modifying lighting, selecting optimal viewpoints, and applying 2D stylization effects. 
An example is illustrated in Figure~\ref{fig:teaser}.
}

At the core of NLI4VolVis is a robust multi-agent architecture where specialized agents collaboratively interpret user intents, generate declarative commands, and iteratively refine visualizations to achieve desired outcomes.
Building on the {\em visualization-perception-action} (VPA) loop introduced by AVA~\cite{Liu-CGF24}, LLM agents iteratively refine visualizations by perceiving semantic details, evaluating progress toward user-defined goals, and taking subsequent actions until the target is achieved. 
The system also employs a flexible semantic segmentation pipeline, accommodating multiple segmentation strategies to ensure robust performance across diverse datasets. 
By integrating these components into an intuitive, interactive interface, NLI4VolVis provides a comprehensive solution for natural language-driven VolVis.
The main contributions of NLI4VolVis are as follows: 
\begin{myitemize} 
\vspace{-0.025in} 
%\item {\bf Comprehensive NLI framework for VolVis}:\ We introduce an NLI framework for VolVis, integrating editable 3DGS to enable users to manipulate large and complex 3D volumetric scenes through natural language querying.
\item \hot{{\bf Semantic-aware real-time rendering and editing for VolVis}:\ We leverage multi-view semantic segmentation to identify scene components and train editable 3DGS models, enabling efficient real-time rendering, editing, and manipulation of volumetric scenes through natural language querying.}

\item {\bf Robust scene understanding pipeline}:\ We develop a scene understanding pipeline that combines vision-language embeddings with entropy-guided multi-view image selection, effectively aligning user queries with target volumetric structures.

\item {\bf Hierarchical multi-agent system powered by LLMs}:\ We design a multi-agent architecture where individual agents follow the VPA loop and collaborate in interpreting user intentions and executing the corresponding visualization tasks.

\item {\bf \hot{Comprehensive} NLI interface}:\ We provide a real-time, interactive interface that integrates visualization and scene editing, voice-enabled communication, transparent agent action logs, and 2D stylization, ensuring an intuitive and seamless user experience.
\vspace{-0.075in} 
\end{myitemize}

%% file: related.tex
\vspace{-0.05in}
\section{Related Work}

\hot{Traditional VolVis methods, such as DVR, offer precise results but are computationally intensive. 
They often require expert knowledge to fine-tune TFs and lack intuitive, semantic-level interaction. 
With the advent of LLMs and vision-language models, NLI has emerged as a promising approach for exploring and manipulating volumetric data more intuitively. 
Building on these advancements, our work integrates a multi-agent LLM framework with editable 3DGS. 
As summarized in Table~\ref{tab:contribution_matrix}, NLI4VolVis unifies five key capabilities:
(1) NLI for visualization,
(2) open-vocabulary querying,
(3) real-time scene editing,
(4) semantic scene understanding, and
(5) user-controlled modifications.
Together, these capabilities significantly improve the flexibility, accessibility, and usability of VolVis systems.}

{\bf Natural language interaction for visualization.}
NLI for visualization has gained significant attention, enabling users to create and interact with visual representations of data using plain language. 
Incorporating well-structured dialogues and contextual understanding greatly improves user interaction in visualization systems~\cite{Voigt-NAACL22}. 
An analysis of visualization-oriented NLI systems highlights the need to resolve query interpretation, data transformation, visual mapping, interaction, and presentation issues to enhance the overall user experience~\cite{Shen-VIS23}.
%Recent surveys have identified key trends and challenges in this evolving research area. 
%Voigt et al.\ \cite{Voigt-NAACL22} emphasized the importance of dialogue structures and contextual understanding to enhance user interaction. 
%Shen et al.\ \cite{Shen-VIS23} provided a comprehensive overview of advancements and challenges in visualization-oriented NLI systems from the perspective of visualization pipelines.
%
%Researchers have proposed various NLI systems tailored to specific visualization domains to address these challenges.

Researchers have designed various NLI-driven approaches for specific visualization domains to tackle these challenges. 
For example, FlowSense~\cite{Yu-VIS20} introduces an NLI designed for dataflow visualization systems, enabling users to construct and modify dataflow diagrams using natural language. 
FlowNL~\cite{Huang-VIS22} advances NLI for flow visualization by developing a declarative grammar that translates user queries into visualizations of flow structures. 
It simplifies complex interactions while supporting domain-specific tasks like flow structure derivation and visualization.
%
%In NL2Vis, the automated generation of data visualizations from natural language has been explored extensively. 
%The potential of LLMs for this task has been investigated, revealing the benefits of in-context learning and iterative refinement for improving accuracy \cite{Wu-PACMMOD24}.
Meanwhile, other approaches harness LLMs for automated visualization generation, employing in-context learning and iterative refinement to enhance the accuracy of translating natural language into visual representations~\cite{Wu-PACMMOD24}.
Similarly, VOICE~\cite{Jia-arXiv23} integrates the conversational capabilities of LLMs with visualization, enabling real-time navigation and manipulation of 3D models through natural language commands. 
Another work, T2TF~\cite{Jeong-VISSP24}, focuses on volume rendering and employs a vision-language model to generate TFs that align rendered volumes with user-defined textual descriptions. 
%\st{These works illustrate the potential of leveraging LLMs and vision-language models to bridge the gap between textual input and complex visualization tasks.}

While these works in SciVis lay the foundation for natural language-driven visualization, they fall short when handling complex semantics in 3D volume data. 
In contrast, our NLI4VolVis extends the NLI paradigm to VolVis by integrating LLM agents with editable 3DGS. 
By incorporating open-vocabulary querying through a vision-language model, NLI4VolVis empowers users to interactively manipulate any object within a 3D scene using natural language. 
This includes adjusting color, opacity, and lighting, selecting the optimal view, and applying style transfer.

% This paragraph is not strongly related to my work
% Building on the NL2Vis paradigm, Luo et al.\ \cite{Luo-VIS2022} proposed a Transformer-based sequence-to-sequence model named ncNet to directly translate natural language queries into visualizations. With optimizations tailored to visualization tasks, ncNet achieves robust performance on standard benchmarks​. Complementary to such neural approaches, NL4DV \cite{Narechania-VIS2021} provides a toolkit for generating visualization specifications from natural language, enabling developers to incorporate natural language capabilities into their systems with minimal NLP knowledge. This modular approach simplifies the implementation of V-NLI systems.

\vspace{-.1in} 
\begin{table}[h]
	\caption{\hot{Comparison of related works on NLI for visualization.}}
	\label{tab:contribution_matrix}
	\vspace{-0.1in}
	\centering
	\resizebox{\columnwidth}{!}{
	\begin{tabular}{c|ccccc}
		%\toprule
		%work & 
		%\multicolumn{1}{m{2.5cm}}{\centering \shortstack{NLI for \\ visualization}} & 
		%\multicolumn{1}{m{2.5cm}}{\centering \shortstack{open-\\vocabulary \\ querying}} & 
		%\multicolumn{1}{m{2.5cm}}{\centering \shortstack{real-time \\ scene \\ Editing}} & 
		%\multicolumn{1}{m{2.5cm}}{\centering \shortstack{semantic \\ scene \\ understanding}} & 
		%\multicolumn{1}{m{2.5cm}}{\centering \shortstack{user-controlled \\ modification}} \\ \hline
		%\midrule
					& NLI		& open-		& real-time	& semantic & user- \\
					& for			& vocabulary	& scene		& scene	  & controlled \\
		method		& visualization 	& querying	& editing		& understanding	& modification \\ \hline
		FlowSense~\cite{Yu-VIS20}       & \ding{51} & \ding{55} & \ding{55} & \ding{55} & \ding{55} \\
		FlowNL~\cite{Huang-VIS22}       & \ding{51} & \ding{55} & \ding{55} & \ding{55} & \ding{51} \\
		VOICE~\cite{Jia-arXiv23}        & \ding{51} & \ding{55} & \ding{55} & \ding{55} & \ding{51} \\
		T2TF~\cite{Jeong-VISSP24}       & \ding{55} & \ding{55} & \ding{55} & \ding{51} & \ding{51} \\
		iVR-GS~\cite{Tang-PVIS25}       & \ding{55} & \ding{55} & \ding{51} & \ding{55} & \ding{51} \\
		LangSplat~\cite{Qin-CVPR24}     & \ding{55} & \ding{51} & \ding{55} & \ding{51} & \ding{55} \\
		AVA~\cite{Liu-CGF24}            & \ding{51} & \ding{55} & \ding{55} & \ding{51} & \ding{51} \\
		ChatVis~\cite{Mallick-ICHPC24}  & \ding{51} & \ding{55} & \ding{55} & \ding{55} & \ding{51} \\
		ParaView-MCP~\cite{Liu-arXiv25}  & \ding{51} & \ding{55} & \ding{51} & \ding{51} & \ding{51} \\
		NLI4VolVis (ours)      		& \ding{51} & \ding{51} & \ding{51} & \ding{51} & \ding{51} \\
		%\bottomrule
	\end{tabular}
}
\end{table}

\begin{figure*}[htb]
	\centering
	\includegraphics[width=0.95\linewidth]{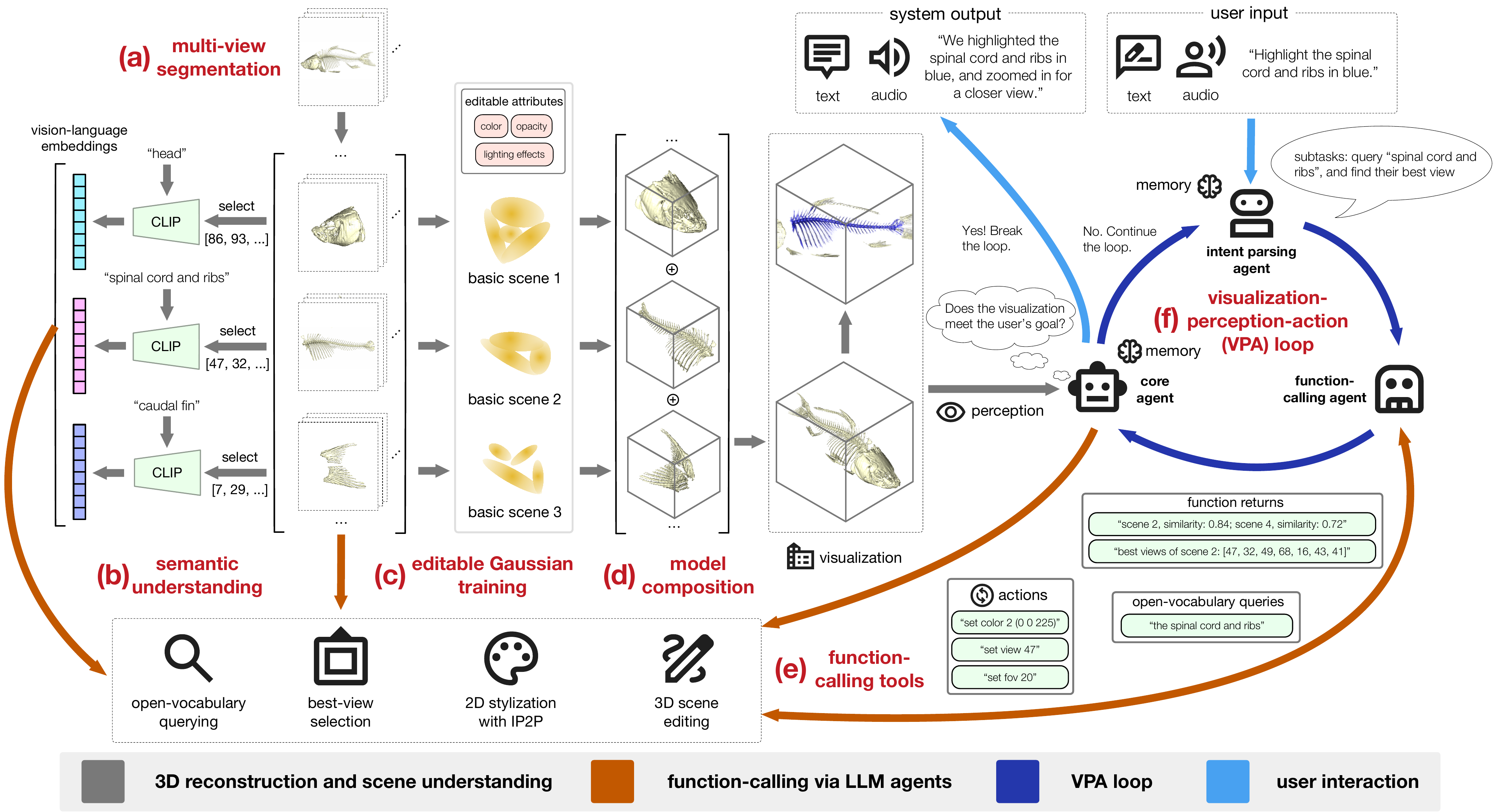}
	\vspace{-.1in}
	\caption{\hot{The NLI4VolVis pipeline. 
			(a) Multi-view segmentation supports cross-frame tracking and 3DGS-based techniques for datasets that are not previously segmented. 
			(b) Each segmented component undergoes entropy-guided view selection, embedding informative frames and textual descriptions with CLIP for semantic understanding.
			(c) Editable iVR-GS models are trained per component. 
			(d) Individual iVR-GS models are composed into the full volumetric scene. 
			(e) Function-calling tools allow LLM agents to query, analyze, edit, and stylize the scene. 
			(f) LLM agents iteratively refine visualizations via the VPA loop based on user input.}}
	\label{fig:pipeline}
\end{figure*}

{\bf Novel view synthesis.}
\begin{comment}
Recent NVS techniques have significantly enhanced 3D scene reconstruction from multi-view images. 
NeRF~\cite{Mildenhall-ECCV20} pioneered this direction, providing high-fidelity geometric modeling but facing challenges in training and inference costs. 
Subsequent methods, such as Instant-NGP~\cite{Muller-TOG22} and Plenoxels~\cite{Keil-CVPR22}, have improved efficiency, making NVS more practical for a broader range of applications. 
In DL4SciVis~\cite{Wang-TVCG23}, NVS belongs to visualization generation~\cite{Berger-TVCG18, Hong-PVIS19, He-InsituNet, Han-TVCG23}, going beyond data generation~\cite{Gu-CGA21, Han-CG22, Han-VI22, Yao-CG23, Gu-PVIS22, Tang-CG24} and neural compression~\cite{Tang-PVIS24, Gu-CG23, YF-Lu-VISSP24, Yang-PVISVN25}. 
In VolVis, methods like StyleRF-VolVis~\cite{Tang-VIS24} and ViSNeRF~\cite{Yao-PVIS25} have extended NVS to VolVis, with the former focusing on style transfer for volume rendering and the latter leveraging a multidimensional radiance field for dynamic scene exploration.
\end{comment}
Recent NVS techniques have significantly enhanced 3D scene reconstruction from multi-view images. 
NeRF~\cite{Mildenhall-ECCV20} pioneered this direction, providing high-fidelity geometric modeling but facing challenges in training and inference costs. 
%NeRF~\cite{Mildenhall-ECCV20} sets the foundation with high-fidelity modeling, though it is limited by high training and inference costs. 
Efficiency-focused methods like Instant-NGP~\cite{Muller-TOG22} and Plenoxels~\cite{Keil-CVPR22} have since broadened NVS applicability. 
In DL4SciVis~\cite{Wang-TVCG23}, NVS is categorized under visualization generation~\cite{Berger-TVCG18, Hong-PVIS19, He-InsituNet, Han-TVCG23}, distinct from data generation~\cite{Gu-CGA21, Han-CG22, Han-VI22, Yao-CG23, Gu-PVIS22, Tang-CG24} and neural compression~\cite{Tang-PVIS24, Gu-CG23, YF-Lu-VISSP24, Yang-PVISVN25}. 
In VolVis, methods like StyleRF-VolVis~\cite{Tang-VIS24} and ViSNeRF~\cite{Yao-PVIS25} have extended NVS to VolVis, with the former focusing on style transfer for volume rendering and the latter leveraging a multidimensional radiance field for dynamic scene exploration.

\begin{comment}
To address the limitations of traditional NeRF methods, 3DGS~\cite{Kerbl-TOG23} introduces explicit Gaussian primitives for real-time rendering and efficient scene composition. 
The application of 3DGS in cinematic anatomy has demonstrated photorealistic rendering of high-resolution anatomical datasets on consumer-grade devices \cite{Simon-arXiv23}.
iVR-GS~\cite{Tang-PVIS25} further enhances Gaussian primitives to support real-time editing of scene attributes, including color, opacity, and lighting. 
It also inspires subsequent works such as TexGS-VolVis~\cite{Tang-VIS25} that uses 2DGS for volume stylization and VolSegGS~\cite{Yao-VIS25} that uses deformable 3DGS for dynamic VolVis scene segmentation and tracking. 
%\st{iVR-GS also supports the composition of models optimized on disjoint TFs, enabling comprehensive exploration of VolVis datasets. Additionally, iVR-GS employs the Blinn-Phong reflection model for precise relighting and offers inverse volume exploration capabilities, making it a robust solution for interactive and explorable VolVis.} 
Given its ability to integrate real-time rendering with interactive editing and scene composability, iVR-GS is an ideal foundation for NLI4VolVis. 
It allows users to intuitively modify volumetric scene properties—such as color, opacity, and lighting—using natural language querying, making it an effective solution for interactive and explorable VolVis.
\end{comment}
To overcome NeRF's limitations, 3DGS~\cite{Kerbl-TOG23} introduces explicit Gaussian primitives for real-time rendering and efficient scene composition. 
Its application in cinematic anatomy~\cite{Simon-arXiv23} showcases photorealistic rendering of high-resolution datasets on consumer devices. 
iVR-GS~\cite{Tang-PVIS25} extends 3DGS to enable real-time editing of scene attributes such as color, opacity, and lighting.
It inspires works such as TexGS-VolVis~\cite{Tang-VIS25} that uses 2DGS for volume stylization and VolSegGS~\cite{Yao-VIS25} that uses deformable 3DGS for dynamic VolVis scene segmentation and tracking. 
With its support for real-time rendering and intuitive editing, iVR-GS provides a strong foundation for NLI4VolVis, enabling NLI for interactive and explorable VolVis.

\hot{StyleGaussian~\cite{Liu-arXiv24} extends 3DGS to enable instant zero-shot 3D style transfer with strong multi-view consistency. 
However, as demonstrated in Section 3 of the appendix, applying StyleGaussian to SciVis data introduces noticeable artifacts and requires substantial training time for each scene. 
Moreover, like most 3D style transfer methods, StyleGaussian alters only the scene's color attributes while preserving the underlying geometry, limiting its expressiveness for more creative or structural editing tasks. 
In contrast, we adopt a 2D stylization approach using InstructPix2Pix (IP2P)~\cite{Brooks-CVPR23}, which supports flexible, prompt-driven creative editing of individual scene components. 
While this approach has known limitations in maintaining multi-view consistency, it offers greater stylistic versatility and immediate control. 
We plan to incorporate more advanced 3D stylization techniques in future iterations of the system to further enhance its capabilities.}

{\bf 3D open-vocabulary segmentation.}
3D open-vocabulary segmentation allows users to interact with any object in a 3D scene using natural language querying, making it a cornerstone of our work. 
Unlike closed-set methods~\cite{Cheng-CVPR22, Zhi-CVPR21}, %Siddiqui-CVPR23}, 
which are limited by predefined categories, open-vocabulary approaches leverage 2D foundational vision-language models (e.g., CLIP~\cite{Radford-ICML21} and SAM~\cite{Kirillov-ICCV23}) to integrate semantic understanding into 3D representations, enabling the recognition and manipulation of arbitrary objects. 
A common strategy is to map 2D semantic features back to 3D space. 
Examples like LangSplat~\cite{Qin-CVPR24} and LERF~\cite{Kerr-ICCV23} embed language-encoded features from 2D multi-view images into 3D fields, enabling flexible querying. 
Advanced methods, such as Semantic Gaussians~\cite{Guo-arXiv24-1} and SAGD~\cite{Hu-arXiv25}, improve segmentation accuracy by incorporating radiance and semantic information using 3DGS.
\hot{OpenGaussian\cite{Wu-NeurIPS24} and Gaussian Grouping\cite{Ye-ECCV24} further this direction by enriching 3D Gaussian primitives with semantic encodings or identity embeddings, enabling point-level and instance-level segmentation and editing. 
These methods facilitate fine-grained 3D scene understanding and manipulation by extending 2D vision-language representations into 3D space.}
NLI4VolVis adopts a flexible pipeline that supports a range of open-vocabulary segmentation techniques (e.g., LangSplat~\cite{Qin-CVPR24}, SAM 2~\cite{Ravi-arXiv24}, and SAGD~\cite{Hu-arXiv25}), as well as pre-segmented datasets, establishing a robust framework for NLI in VolVis.

{\bf LLM-based autonomous agents.}
LLM-based autonomous agents have emerged as a powerful solution for human-like decision-making across diverse domains, driving significant advancements in their development, applications, and evaluation~\cite{Wang-FCS24, Guo-arXiv24-2, Tran-arXiv25}. 
%LLM-based autonomous agents have emerged as a powerful solution for achieving human-like decision-making capabilities across diverse domains. 
These agents are designed to dynamically plan, reason, and interact with their environment by leveraging LLMs' vast internal world knowledge. 
%Recent surveys~\cite{Wang-FCS24, Guo-arXiv24-2, Tran-arXiv25} offer a comprehensive overview of the development, applications, and evaluation of LLM-based autonomous agents, highlighting their adaptability and versatility.
%
In visualization, such agents have been applied to automate data exploration and visualization workflows. 
For instance, AVA~\cite{Liu-CGF24} combines natural language understanding with visual perception to iteratively refine visualizations based on user-defined objectives, such as adjusting TFs or hyperparameters in volume rendering. 
Similarly, ChatVis~\cite{Mallick-ICHPC24} automates SciVis processes by generating and iteratively refining Python scripts for tools like ParaView, enabling non-experts to create visualizations using natural language.
\hot{ParaView-MCP~\cite{Liu-arXiv25} integrates LLM agents with direct API control in ParaView through the {\em model context protocol} (MCP), enabling seamless information exchange among users, LLM agents, and visualization tools to support human-AI collaborative visualization workflows.}
NLI4VolVis builds upon these advances by integrating LLM-based agents into the NLI pipeline to interpret user intent and execute VolVis scene manipulations. 
These agents facilitate interaction with 3D scenes through natural language, supporting open-vocabulary querying, interactive editing, and exploratory tasks. 
By harnessing LLMs' planning and reasoning capabilities, we make VolVis more accessible across various scenarios.

%% file: method.tex
\vspace{-0.05in}
\section{Our Approach}

As illustrated in Figure~\ref{fig:pipeline}, our NLI4VolVis framework offers a comprehensive pipeline that integrates multi-view semantic segmentation, 3D open-vocabulary scene understanding, editable Gaussian training, and NLI for VolVis. 
The pipeline begins with multi-view image segmentation of the volumetric scene using flexible semantic segmentation strategies (Section~\ref{subsec:mvss}). 
Each segmented object is then processed through an entropy-guided view selection mechanism, where the most informative frames are embedded using the CLIP model to achieve robust open-vocabulary scene understanding (Section~\ref{subsec:3dov}). 
This ensures that user queries are accurately aligned with the corresponding volumetric structures.
Next, we train editable NVS models for each segmented component as a basic scene using iVR-GS~\cite{Tang-PVIS25}, enabling independent editing and seamless composition of the complete scene.
The resulting models support real-time editing of color, opacity, and lighting, as well as composability through the concatenation of Gaussian parameters (Section~\ref{subsec:egt}). We introduce a hierarchical multi-agent system powered by LLMs to enable natural language-driven interaction. 
LLM agents utilize external function-calling tools to interpret and modify 3D scenes, enabling open-vocabulary querying, optimal view selection, interactive scene editing, and 2D stylization.
These collaborative agents follow the VPA loop~\cite{Liu-CGF24}, allowing them to interpret user intents, generate declarative commands, and iteratively refine visualizations (Section~\ref{subsec:NLI4VolVis}).
Finally, our interactive user interface (Section~\ref{subsection:ui}) integrates all components of the pipeline, enabling  NLI with volume datasets through features such as real-time visualization editing, transparent agent action logs, speech input and audio output, and 2D stylization via IP2P~\cite{Brooks-CVPR23}.

\hot{When a new volumetric dataset is introduced, we begin by rendering multi-view images using a preset TF in ParaView. 
These images are then processed by a selected multi-view semantic segmentation method. 
Each segmented semantic component is used to train an independent, editable 3DGS model. 
The final scene is assembled by combining these component models, enabling real-time editing, open-vocabulary querying, and stylization within NLI4VolVis.}

\vspace{-0.05in}
\subsection{Multi-View Semantic Segmentation}
\label{subsec:mvss}

To enable open-vocabulary querying and interactive scene editing in VolVis, our NLI4VolVis pipeline must support efficient semantic segmentation across various VolVis scenarios. 
Since an NVS model can generate new viewpoints from multi-view images, a natural approach is to train editable 3DGS models using segmented multi-view images. 
In this setup, each segmented object corresponds to an independent iVR-GS model~\cite{Tang-PVIS25}, ensuring that individual objects can be queried and edited separately.

The first step in this pipeline is multi-view semantic segmentation. 
Given the diversity of VolVis datasets, a one-size-fits-all segmentation method is impractical. 
Therefore, we adopt a flexible approach, allowing users to select the segmentation method that best suits their data. 
Below, we summarize commonly used semantic segmentation techniques in VolVis, discussing their strengths and limitations.

{\bf Pre-segmented datasets.} 
Some volumetric datasets, such as medical imaging benchmarks (e.g., FLARE human organ~\cite{FLARE22-LDH24}), come with ground-truth semantic labels. 
These datasets are ideal for controlled environments but do not generalize well to arbitrary VolVis scenes.

{\bf TF-based segmentation.} 
Traditional TFs in DVR map voxel values to color and opacity. 
However, they are not well-suited for intuitive, object-level segmentation, particularly for non-expert users. 
Even multi-dimensional TFs often fail to segment objects based on semantic meaning~\cite{Ljung-CGF16}, as they primarily rely on voxel intensity and derived attributes rather than learned object semantics.

{\bf Voxel-based segmentation.} 
Voxel-based techniques extract features using voxel coordinates, scalar values, gradients, and neighborhood information~\cite{Poux-ISPRS19}. 
Approaches like voxel2vec~\cite{He-VIS23} apply representation learning to uncover semantic structures within a volumetric dataset. 
While voxel-based methods effectively cluster local regions, they often lack semantic awareness and struggle with generalization to complex, real-world 3D scenes. 
Additionally, since NVS models operate on multi-view images rather than raw voxels, directly applying voxel-based segmentation is not well-suited to our pipeline.

%{\bf Framewise 2D segmentation.} 
%\hot{Turn this into the motivation for cross-frame object tracking}. A straightforward approach is to apply 2D foundation models, such as SAM~\cite{Kirillov-ICCV23}, to each rendered viewpoint separately. 
%However, this can lead to view inconsistency, where an object's mask varies between neighboring frames due to occlusions, lighting changes, or segmentation artifacts.

{\bf Cross-frame object tracking.} 
A straightforward approach to segmenting multi-view images is to apply 2D foundation models, such as SAM~\cite{Kirillov-ICCV23}, to each rendered viewpoint independently. 
However, this framewise segmentation often results in view inconsistencies, where an object's mask varies between neighboring frames due to occlusions, lighting changes, or segmentation artifacts.
To address this, we adopt SAM 2~\cite{Ravi-arXiv24}, a video object segmentation and tracking model that improves segmentation consistency across viewpoints. 
By sampling along a continuous circular camera trajectory, we treat multi-view images as a video sequence, enabling objects to be tracked across frames and reducing inconsistencies caused by independent framewise segmentation. 
However, SAM 2 struggles with transparent and nested objects, which are common in VolVis.

{\bf Direct segmentation in 3DGS.} 
These methods perform semantic segmentation directly on NVS models by leveraging 2D foundation models to integrate semantic features into 3D Gaussian representations. 
Since segmentation is applied directly to the 3D Gaussians, segmented renderings can be generated from any viewpoint, ensuring view-consistent object querying. 
However, this approach modifies the underlying Gaussian structure, reducing flexibility and making it less compatible with other segmentation methods. 
To ensure scalability and interoperability, NLI4VolVis does not directly segment iVR-Gaussians. 
Instead, it trains iVR-GS models using multi-view renderings of objects segmented by LangSplat~\cite{Qin-CVPR24} and SAGD~\cite{Hu-arXiv25}. 
While this method maintains a method-agnostic segmentation approach, it may result in some loss of visualization quality.

In summary, each segmentation method has its own strengths and trade-offs. 
Given the diverse nature of VolVis scenes, no single method is optimal for all use cases. 
By leveraging segmented multi-view images as input to editable 3DGS models, NLI4VolVis adopts a flexible strategy that allows users to choose the segmentation method best suited to their specific datasets and visualization goals.

\vspace{-0.05in}
\subsection{3D Open-Vocabulary Scene Understanding via CLIP}
\label{subsec:3dov}

To facilitate natural language-driven VolVis scene editing, NLI4VolVis integrates 3D open-vocabulary scene understanding using the CLIP model~\cite{Radford-ICML21}. 
This integration enables the identification of objects within a 3D scene based on user queries expressed in natural language.

Although CLIP is inherently a 2D model, we aim to generate embeddings for 3D objects. 
Since different viewpoints of the same 3D object reveal varying information, selecting representative views is crucial. 
We employ an entropy-based method~\cite{Ji-TVCG06} to choose the top-$k$ frames from evenly sampled multi-view images of a segmented object. 
Each selected frame is processed to obtain its CLIP embedding, and the mean of these embeddings serves as the final representation.

To identify the most informative viewpoints, we calculate the entropy of each image based on its opacity (alpha) channel. 
Using opacity alone allows us to assess the structural complexity within the volume data, as it is less affected by lighting and shading effects compared to color. 
Including color in the calculation may introduce variability, leading to a less consistent representation of the underlying structure. 
Thus, we focus solely on opacity to ensure a more reliable measure. 
Images with higher entropy contain more information and are, therefore, more informative. 
The entropy $H$ of an image $I$ is computed as
	$H(I)=-\sum_{i=1}^{N} p_{i}\log p_{i}$,
where $p_{i}$ denotes the probability (normalized opacity value) at pixel $i$, and $N$ is the total number of pixels. 
After computing the entropy for all frames, we select the top-$k$ frames with the highest entropy values.

For each of the selected top-$k$ frames, we generate a CLIP embedding. 
The vision-based embedding $E_{\text{vision}}$ for the 3D object is obtained by averaging these individual embeddings
	$E_{\text{vision}}=(\sum_{j=1}^{k} E_{\text{frame}_{j}})/k$,
where $E_{\text{frame}_{j}}$ denotes the CLIP embedding of the $j$-th selected frame. 
For datasets without additional textual information, the object embedding is defined solely by the vision-based embedding, i.e., 
$E_{\text{object}} = E_{\text{vision}}$.
When textual labels or descriptions are available, we also utilize CLIP to embed the corresponding textual description. 
The final object embedding is then computed as
$E_{\text{object}} = (E_{\text{vision}} + E_{\text{text}}) / 2$,%\end{equation}​
where $E_{\text{text}}$ is the text-based embedding. 
This integration of vision and textual embeddings creates a unified, multimodal representation for each semantic component. 
The unified embedding enables more accurate and efficient open-vocabulary querying by capturing visual and semantic cues in a single vector space.

Once the CLIP embeddings are established for each segmented object, our system can interpret user queries expressed in natural language. 
Upon receiving a query, we generate its CLIP embedding and compute $S(E_{\text{query}} ,\ E_{\text{object}_{i}})$, the cosine similarity between this query embedding $E_{\text{query}}$ and the embedding of each object $E_{\text{object}_{i}}$
\begin{equation}
	S(E_{\text{query}} ,\ E_{\text{object}_{i}}) =\frac{E_{\text{query}} \cdot E_{\text{object}_{i}}}{\| E_{\text{query}} \| \cdot \| E_{\text{object}_{i}} \| }.
\vspace{-0.05in}
\end{equation}
The objects corresponding to the highest similarity scores are identified as the target of the user's query, enabling efficient scene editing based on natural language input.

\vspace{-0.05in}
\subsection{Editable Gaussian Training}
\label{subsec:egt}

Existing NVS methods for VolVis~\cite{Tang-VIS24, Yao-PVIS25, Tang-PVIS25, Yao-CG25} cannot semantically locate and edit objects. 
One promising solution is to apply semantic segmentation techniques (Section~\ref{subsec:mvss}) to partition VolVis scene components, representing each as a set of multi-view images. 
These segmented multi-view images are then used as input to train an NVS model for each individual component. 
The {\em editability} of the NVS model ensures that each reconstructed 3D component remains editable. 
Ideally, the NVS method should also support {\em composability}, meaning several independently trained models, referred to as basic scenes, can be combined by simply concatenating their corresponding parameters. 
This feature allows users to view individual semantic components separately while still providing a complete reconstruction of the original VolVis scene.

We select iVR-GS~\cite{Tang-PVIS25} as the NVS method for our work. 
iVR-GS builds upon 3DGS~\cite{Kerbl-TOG23}, an explicit 3D scene representation technique. 
Each 3D Gaussian primitive $G(\mathbf{x})$ is characterized by its spatial mean $\mathbf{\mu} \in \mathbb{R}^{3 \times 1}$ and covariance matrix $\Sigma \in \mathbb{R}^{3 \times 3}$
\begin{equation}
	G(\mathbf{x}) =\exp(-\frac{1}{2}(\mathbf{x}-\mathbf{\mu})^\top \Sigma^{-1}(\mathbf{x} -\mathbf{\mu})),
\vspace{-0.05in}
\end{equation}
where $\mathbf{x}\in\mathbb{R}^{3 \times 1}$ represents a spatial coordinate. 
Each original 3D Gaussian also has two additional attributes: an opacity value $o\in [0,1]$ and a view-dependent color $\mathbf{c} \in \mathbb{R}^{3 \times 1}$. 
For differentiable optimization, the covariance matrix $\Sigma$ is further parameterized by a scaling vector $\mathbf{s} \in \mathbb{R}^{3 \times 1}$ and a rotation vector $\mathbf{q} \in \mathbb{R}^{4 \times 1}$. 
3DGS uses a set of 3D Gaussians to reconstruct a 3D scene, and the $i$-th Gaussian primitive is parameterized with $\{\mathbf{\mu}_{i} ,\mathbf{s}_{i} ,\mathbf{q}_{i} ,o_{i} ,\mathbf{c}_{i}\}$.

3D Gaussians can be projected onto the image plane through rasterization when rendering from a novel view. 
With the viewing transformation matrix $\mathbf{W}$ and the Jacobian matrix $\mathbf{J}$ of the projective transformation, the projected 2D covariance matrix $\Sigma^{\prime}$ is computed as $\Sigma^{\prime} =\mathbf{J} \mathbf{W} \Sigma \mathbf{W}^\top \mathbf{J}^\top$. 
For each pixel, its color $\mathbf{C}$ is determined by the sum of $N$ sequentially layered 2D Gaussians alone the ray
\begin{equation}
	\mathbf{C} =\sum_{i\in N} \mathbf{c}_{i} \alpha_{i} \prod_{j=1}^{i-1}(1-\alpha_{j}),
\vspace{-0.05in}
\end{equation}
where $\alpha$ is the alpha value computed by multiplying the Gaussian weight with the opacity $o$ learned per-primitive. 
iVR-GS extends the original 3DGS by augmenting primitives with extra attributes and optimizing them to formulate editable Gaussian primitives. 

The training of an iVR-GS model starts from optimizing an original 3DGS model with an extra normal attribute $\mathbf{n}$ for each Gaussian. 
This normal is used to calculate the view-dependent color $\mathbf{c}$ for each Gaussian based on the Blinn-Phong reflection model. 
To enable color and lighting edits, several extra attributes are assigned to each Gaussian in iVR-GS: an offset color $\Delta \mathbf{c} \in \mathbb{R}^{3 \times 1}$, ambient, diffuse, and specular coefficients $\{k_{a}, k_{d}, k_{s}\}$, and a shininess term $\beta$. 
Specifically, the voxel color in the Blinn-Phong reflection model is expressed as $\mathbf{c}_{p} +\Delta \mathbf{c}$, where the palette color $\mathbf{c}_{p} \in \mathbb{R}^{3 \times 1}$ is frozen during training as the mean color value of all multi-view training images and is shared by all Gaussians within a basic scene. 

In addition to the loss functions used in 3DGS, iVR-GS incorporates two key regularization strategies. 
First, L1 sparsity regularization is applied to the offset color $\Delta \mathbf{c}$ to prevent excessive color shifts. 
Second, bilateral smoothness regularization is applied to the shading attributes $\{k_{a}, k_{d}, k_{s}, \beta\}$, ensuring smooth shading transitions in uniform-color regions. 
After training, the scene is represented by a set of editable Gaussians, where the $i$-th primitive is defined by 
shading attributes $\{k_{ai}, k_{di}, k_{si}, \Delta \mathbf{c}_{i}, \beta_{i}\}$ and 
geometry attributes $\{\mathbf{\mu}_{i}, \mathbf{s}_{i}, \mathbf{q}_{i}, o_{i}\}$.

During inference, these editable Gaussians in iVR-GS allow for the editing of color, opacity, light magnitude, and light direction. 
In particular, we train one basic iVR-GS model for each segmented component in NLI4VolVis. 
To edit the color of each component, we replace its $\mathbf{c}_{p}$ with the user-desired color $\mathbf{c}^{\prime}_{p}$​. 
The opacity of the component is adjusted by scaling the opacity term $o$. 
The coefficients $k_{a}, k_{d}, k_{s}$ and $\beta$ adjust the lighting magnitude accordingly. 
Finally, by using the normal $\mathbf{n}$ of each Gaussian primitive, changes to lighting direction are achieved following the Blinn-Phong equations.

In contrast to 3DGS, iVR-GS also supports composability: combining multiple scenes by simply concatenating Gaussian parameters. 
This composability lets users view each semantic component individually or reconstruct the complete VolVis scene in NLI4VolVis.

\vspace{-0.05in}
\subsection{NLI for VolVis}
\label{subsec:NLI4VolVis}

In the NLI4VolVis framework, we combine editable 3DGS for scene representation with various semantic segmentation techniques to extract and manipulate semantic components within VolVis. 
The NLI component utilizes the extensive world knowledge and visual understanding of multimodal LLMs to interpret user intentions and execute corresponding visualization tasks using a suite of function-calling tools. 
We also develop a set of declarative command rules that define visualization operations—such as adjusting opacity, color, and lighting—for each semantic component in the scene. 
To further enhance system capabilities, we adopt a collaborative multi-agent architecture.

{\bf Multi-agent collaboration.}
Recent advances in LLMs have enabled the development of agentic AI, where multiple LLM-based agents can perceive, reason, and collaborate to solve complex tasks in real-world applications. 
These multi-agent systems consist of intelligent agents that communicate, collaborate, and sometimes compete to address tasks at scale. 
The relationships among agents can be cooperative, competitive, or a combination of both, and collaboration strategies can be role-based, rule-based, or model-based~\cite{Guo-arXiv24-2, Tran-arXiv25}. 
In NLI4VolVis, we implement a rule-based, collaborative multi-agent system with a hierarchical communication structure. 
This system includes agents responsible for tasks such as parsing, open-vocabulary querying, user interaction through dialogue, and declarative command generation. 
To efficiently manage explicit VolVis tasks, NLI4VolVis employs a hierarchical communication structure where agents are assigned distinct roles across multiple layers. 
These agents collaborate iteratively, ensuring that the core agent can determine when the user's intent has been fulfilled or when the maximum number of iterations has been reached. 
This hierarchical structure effectively distributes tasks across different levels, enhancing the robustness and resilience of the reasoning process.

Each agent in the system can observe and perceive its environment, reason based on its observations, and then take action. 
Current LLM agents typically produce output in the form of {\em text}, {\em audio}, or {\em images}. 
Their ability to perceive and manipulate the environment can be significantly enhanced using external {\em function-calling tools}~\cite{Qin-ACMCS24, Guo-ACLFindings24}. 
In NLI4VolVis, each agent has its own memory and a task-driven planning core defined by prompts that describe specific tasks. 
Agents can also access multiple function-calling tools to achieve visualization goals.

Moreover, agents in NLI4VolVis must comprehend the outcomes of their visualizations and adapt their actions accordingly. 
To support this, we incorporate the VPA loop introduced by AVA~\cite{Liu-CGF24}. 
This loop enables the core agent to manipulate semantic components within VolVis by sending declarative commands to the visualization server. 
It also leverages the visual perception capability to understand the current visualization, assess whether the user-defined goal has been achieved, and reason about which declarative commands to issue in the next iteration, if necessary. 
For each user input, the agents will follow this loop iteratively until the core agent determines that the user-defined visualization target has been met.

{\bf Function-calling via LLM agents.}
In NLI4VolVis, LLM agents leverage external function-calling tools to perceive, analyze, and manipulate visualizations in response to user intent. 
These tools offer structured workflows that extend the capabilities of LLMs, allowing them to execute complex tasks beyond their internal reasoning. 
The key function-calling capabilities in NLI4VolVis include
\begin{myitemize}
\vspace{-0.05in}
	\item {\bf Open-vocabulary object querying}:\ Computes CLIP similarity (see Section~\ref{subsec:3dov}) between the user's textual description and precomputed embeddings of segmented components, enabling agents to accurately identify the objects being referred to.
	\item {\bf Scene editing}:\ Allows modifications to each segmented component's color, opacity, and lighting parameters, as well as global adjustments to the VolVis scene.
	\item {\bf Knowledge-based question answering}:\ Extracts dataset-specific metadata to assist agents in understanding the semantic context of the visualization, enhancing their ability to respond to domain-specific user queries.
	\item {\bf Best-view selection}:\ Retrieves the top-$k$ frames with the highest entropy values for each semantic component, allowing agents to adjust the camera view for optimal presentation. \hot{Here, entropy is computed over the alpha channel of the rendered frames; higher entropy indicates richer structural complexity, which helps in selecting more informative viewpoints.}
	\item {\bf 2D stylization}:\ Interfaces with IP2P~\cite{Brooks-CVPR23}, a text-based image diffusion model, enabling agents to apply user-defined styles to specific objects or the entire scene.	
\vspace{-0.05in}
\end{myitemize}
By incorporating these function-calling tools, NLI4VolVis ensures that LLM agents can precisely interpret user intentions and execute the corresponding visualization tasks.

{\bf Declarative VolVis commands.}
The iVR-GS engine in NLI4VolVis operates as a server, receiving declarative commands from LLM agents to dynamically modify the visualization. 
These commands allow precise control over various scene properties and ensure the visualization aligns with the user's intent.
Basic visualization settings include adjusting each segmented object's color, opacity, and lighting parameters. 
Camera controls allow LLM agents to set the viewing direction and field of view, enabling optimal perspectives of specific scene components. 
Additional rendering operations, such as setting the rendering mode, changing the background color, saving images, and resetting views, provide further flexibility.
Moreover, LLM agents utilize declarative commands to apply 2D text-based stylization through IP2P, enabling artistic transformations on selected objects or the entire scene. 
In essence, these agents control every visualization aspect by executing structured commands, ensuring precise modifications that align with user objectives.

\vspace{-0.05in}
\subsection{Interactive Interface}
\label{subsection:ui}

\begin{figure}[htb]
	\centering
		\includegraphics[width=1.0\linewidth]{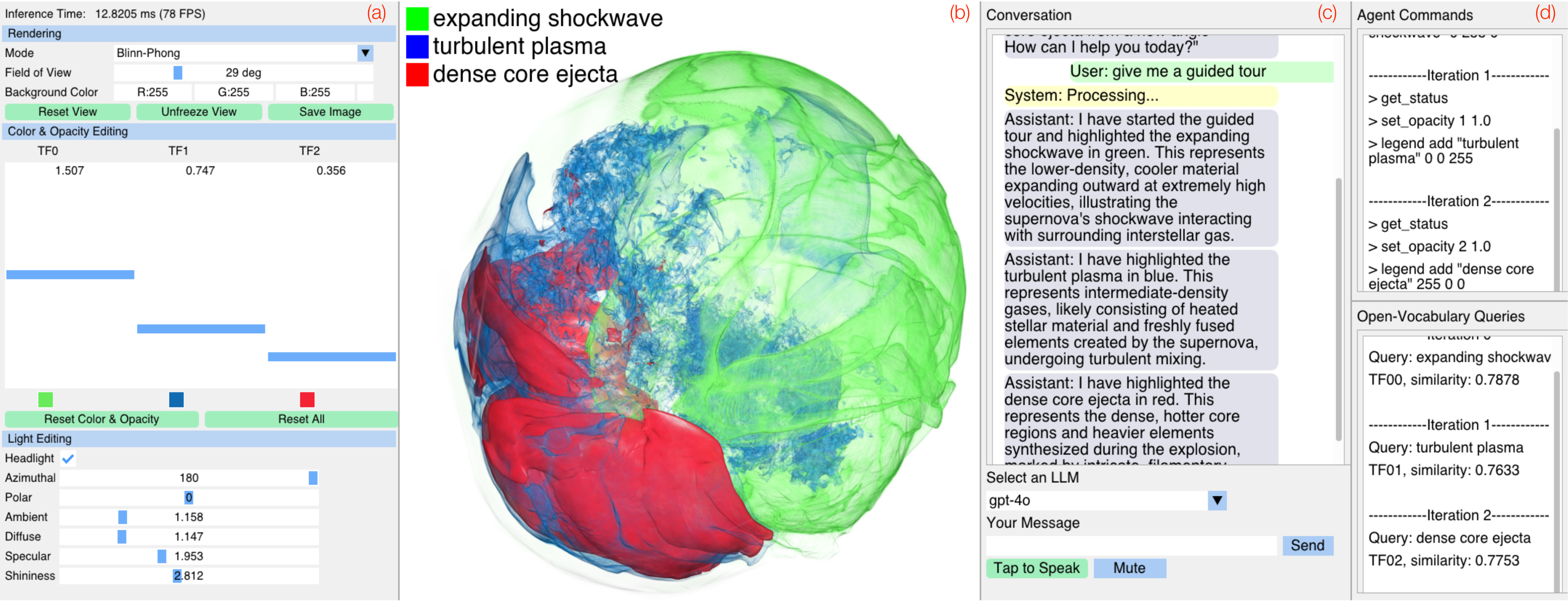}
	\vspace{-0.25in}
	\caption{The NLI4VolVis interface consists of four key components: (a) panel, (b) rendering window, (c) chat widget, and (d) action log.}
%The NLI4VolVis interface. (a) The control panel lets users interactively adjust color, opacity, lighting, and view settings. (b) The rendering window provides real-time visual feedback, dynamically updating based on user commands. (c) The chat widget supports text and voice-based interaction with multi-LLM agents, which interpret user queries, execute visualization edits, and provide guided explanations. (d) The action log maintains a record of all agent-driven declarative commands and open-vocabulary queries, ensuring explainability and facilitating iterative refinement of the visualization. 
	\label{fig:interface}
\end{figure}

The NLI4VolVis interface is designed to facilitate seamless interaction with volumetric data through natural language inputs and direct visualization manipulation. 
As shown in Figure~\ref{fig:interface}, the interface consists of four main components, arranged from left to right: the control panel, rendering window, chat widget, and action log.

\vspace{-0.1in}

The control panel allows users to adjust color, opacity, lighting, and viewing angles to highlight specific structures and refine views. 
The rendering window updates in response to user inputs, reflecting changes made through declarative commands. 
The chat widget enables NLI with LLM agents. 
Users can submit queries via text or speech input and receive responses alongside an updated visualization. 
The widget supports multiple LLMs, including GPT-4o, DeepSeek-V3, and Llama3.2-90B-Vision, allowing users to choose AI models based on their preferences and task requirements.
The action log records all executed commands and open-vocabulary queries, offering transparency into the multi-agent decision-making process. 
This feature helps users understand the reasoning behind visualization changes and refine their queries for more precise results. \hot{In the ``Open-Vocabulary Query'' section of the action log, ``TF'' refers to the transfer function, representing a semantic component segmented using TFs in the 3D scene. ``Similarity'' indicates the cosine similarity between the CLIP embedding of the user query and that of each semantic component, enabling open-vocabulary object identification.}
To further enhance accessibility, NLI4VolVis supports voice-based input and audio responses. 
\hot{Based on user study feedback, we have improved the voice interaction experience and added multilingual support for both input and output.}
Users can also review past interactions in the action log, allowing for iterative refinement of queries to achieve more accurate visualization outcomes. 
For a more detailed demonstration, we refer readers to the supplementary video, which showcases real-time interactions with the system.

%% file: results.tex
%\vspace{-0.05in}
\section{Results and Evaluation}

We evaluate NLI4VolVis across eight diverse volumetric datasets to demonstrate its effectiveness and generalizability. 
As detailed in Table~\ref{tab:dataset}, five datasets were obtained through scanning, encompassing everyday objects, animals, and organs, while the remaining three were generated from scientific simulations. 
\hot{For each dataset, we report the optimal multi-view semantic segmentation method in Table~\ref{tab:dataset-partTFs} and offer clearer guidance on selecting the appropriate method based on volumetric data characteristics in Section 1 of the appendix.} 
Volume sizes range from 50 MB to 4.3 GB, yet 3DGS ensures low-latency inference, consistently achieving over 100 {\em frames per second} (FPS) on an NVIDIA RTX 4090 GPU, regardless of volume size. 
In contrast, DVR with ParaView faces a substantial performance bottleneck, achieving only 5.09 FPS when rendering the largest 4.32 GB dataset, chameleon. 3DGS outperforms DVR by up to 20$\times$. 

\vspace{-0.1in}
\begin{table}[htb]
	\caption{Datasets and their respective settings.}
	\label{tab:dataset}
	\vspace{-0.1in}
	\centering
	%{\scriptsize
		\resizebox{\columnwidth}{!}{
			\begin{tabular}{c|cc|cc}			
				& volume     &  volume &\# semantic   &  segmentation     \\ 
				dataset & resolution 	   &  size	   &  components   &  method \\ \hline
				backpack & 512$\times$512$\times$373 & 373 MB   & 5 & TF+SAGD\\
				carp & 256$\times$256$\times$512 & 64 MB  &7  & SAM 2 \\
				chameleon & 1024$\times$1024$\times$1080 & 4.32 GB  &4  & TF+SAGD\\
				FLARE human organ & 512$\times$512$\times$330 &  330 MB   & 14 & pre-segmentation\\ 
				kingsnake & 1024$\times$1024$\times$795 & 3.18 GB   & 2  & TF\\ \hline
				hurricane &500$\times$500$\times$100 & 95 MB   & 3  & TF \\
				mantle temperature &360$\times$201$\times$180 & 49.69 MB   & 5  & TF \\
				supernova &432$\times$432$\times$432 & 323 MB   & 3  & TF \\
			\end{tabular}
		}
		\label{tab:dataset-partTFs}
	\end{table}

For Gaussian splatting, we set the image resolution to 800$\times$800 for both training and inference. 
The training begins with generating 92 multi-view images using icosphere sampling in ParaView with the NVIDIA IndeX plugin~\cite{ParaViewNVIDIAIndeX}. 
After multi-view semantic segmentation, we train a separate iVR-GS~\cite{Tang-PVIS25} model for each semantic component. 
The training follows a two-stage process: an initial 30,000 iterations for base 3DGS training, followed by 10,000 iterations for editable Gaussians. 
All other parameters remain consistent with the original iVR-GS. 
The average training time for each basic scene is approximately seven minutes.
To highlight the capabilities of NLI4VolVis, we present several illustrative case studies and conduct a formal user study to assess interaction performance. 
Refer to the supplemental video for demonstrations of real-time interactions.
%For multi-view semantic segmentation, we adopt LangSplat~\cite{Qin-CVPR24}, SAGD~\cite{Hu-arXiv25}, SAM 2~\cite{Ravi-arXiv24} and TF-based segmentation.

\vspace{-0.05in}
\begin{figure}[htb]
	\centering
	\includegraphics[width=0.75\linewidth]{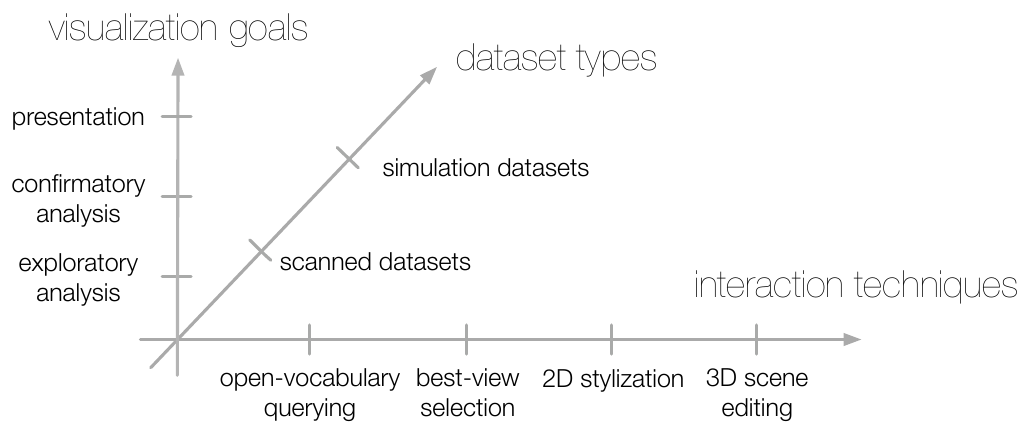}
	\vspace{-0.1in}
	\caption{The three independent axes serve as the foundation for designing our NLI4VolVis case studies.}
	\label{fig:axes}
\end{figure}

\vspace{-0.05in}
\subsection{Case Studies}

%Baumann-ICIV11,
Data visualization serves three primary goals: {\em exploratory analysis}, {\em confirmatory analysis}, and {\em presentation}~\cite{Lam-TVCG18}. 
Exploratory analysis enables open-ended interaction with data, allowing users to uncover patterns and generate hypotheses without predefined assumptions.
Confirmatory analysis involves testing specific hypotheses through structured examination, validating or refuting initial insights.
Presentation focuses on effectively communicating findings by selecting appropriate visual techniques to convey pre-established facts.
NLI4VolVis guides users in achieving these three goals when working with scanned or simulation datasets, leveraging a suite of viewing and editing tools, including open-vocabulary querying, best-view selection, 2D stylization, and 3D scene editing. 
As illustrated in Figure~\ref{fig:axes}, each case study in NLI4VolVis aligns with a subset of these goals along the three independent axes.

{\bf Carp.}
The carp dataset comprises a CT scan of a carp fish, segmented into seven semantic components using SAM 2. 
Each component is processed into a basic scene for further analysis.

\begin{figure}[htb]
 \includegraphics[width=1.0\linewidth]{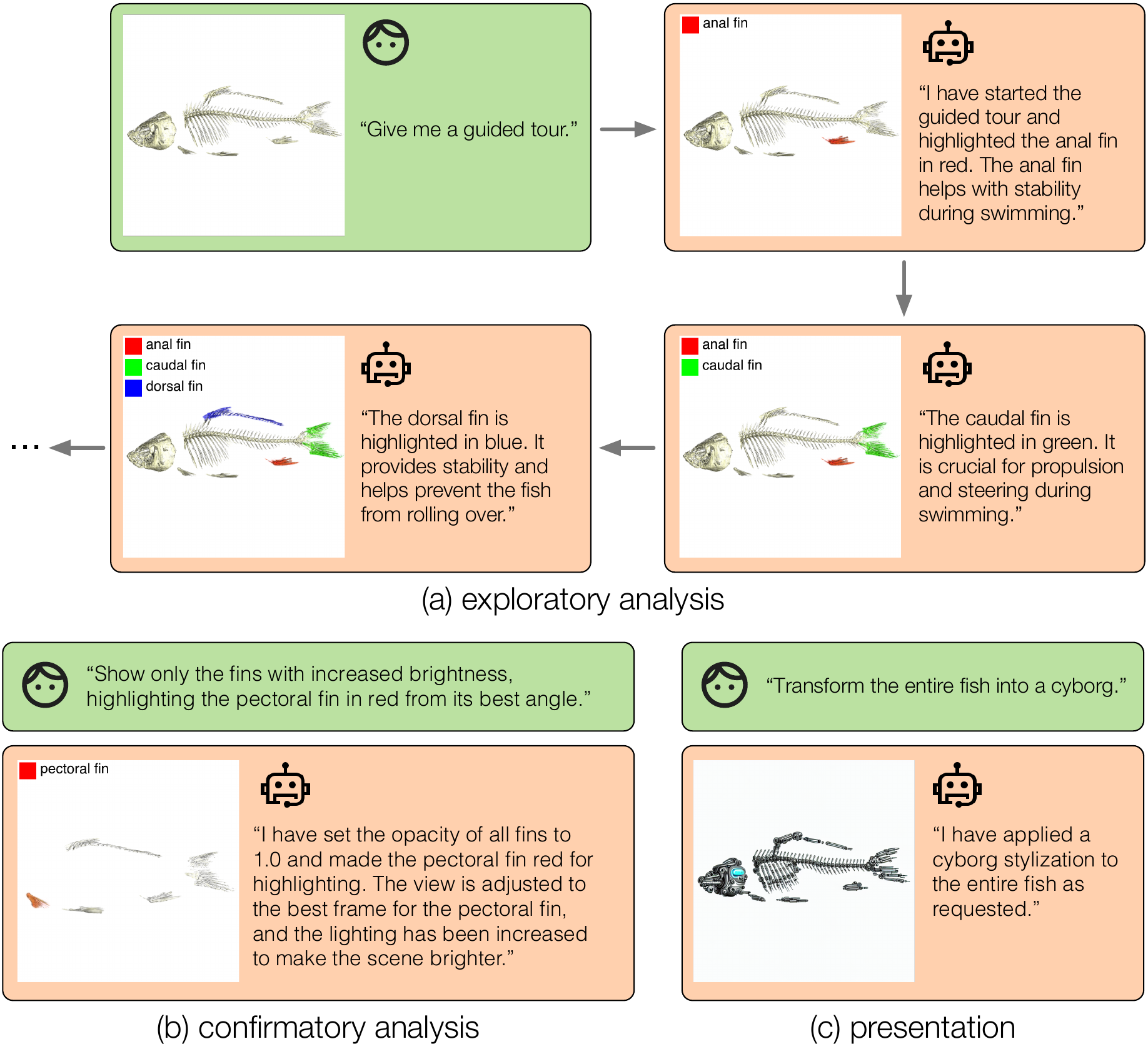}
% \begin{center}
% $\begin{array}{c@{\hspace{0.05in}}c@{\hspace{0.05in}}c}
% \includegraphics[height=1.0in]{figures/carp_1.pdf}&
% \includegraphics[height=1.0in]{figures/carp_2.pdf}&
% \includegraphics[height=1.0in]{figures/carp_3.pdf}\\
% \mbox{\footnotesize (a)} & \mbox{\footnotesize (b)} & \mbox{\footnotesize (c)} 
% \end{array}$
% \end{center}
\vspace{-.25in} 
\caption{Case study with the carp dataset demonstrating the three visualization goals.}
\label{fig:carp}
\end{figure}

To facilitate exploratory analysis, we design a guided tour that introduces users to the dataset by highlighting each semantic component in distinct colors while providing natural language and audio explanations. 
This approach familiarizes users with the dataset, laying the groundwork for deeper analysis. 
As shown in Figure~\ref{fig:carp} (a), NLI4VolVis navigates through the segmented components step by step in response to a simple prompt: ``{\em Give me a guided tour.}''

Following the guided tour, users can proceed with confirmatory analysis by testing specific hypotheses. 
NLI4VolVis supports flexible scene editing, best-view selection, and natural language explanations to help users verify their assumptions. 
For instance, as illustrated in Figure~\ref{fig:carp} (b), users can request the system to show only the fins, highlight the pectoral fin in red, adjust the view for better visibility, and increase brightness. 
NLI4VolVis interprets these requests, identifies the referenced object using open-vocabulary querying, and executes declarative visualization commands while providing explanatory feedback.

For presentation-focused visualization, NLI4VolVis offers extensive 3D scene editing and text-driven 2D stylization to generate artistic representations of the dataset. 
Since 3D stylization methods~\cite{Haque-ICCV23, Vachha-igs2gs24, Chen-DGE} require significant computational time for fine-tuning NVS models, we support real-time 2D stylization instead, enabling immediate visual enhancements. 
Figure~\ref{fig:carp} (c) shows such a result of the carp dataset generated with the prompt: ``{\em Transform the entire fish into a cyborg.}''

A user study was conducted using this dataset to evaluate the effectiveness of NLI4VolVis in data exploration and interaction efficiency. 
See Section~\ref{subsec:us} and the supplementary materials for detailed results.

\vspace{-.1in} 

{\bf FLARE human organ.}
The FLARE human organ~\cite{FLARE22-LDH24} dataset is a labeled abdominal organ dataset from the MICCAI 2022 Grand Challenge, designed for medical imaging and educational purposes. 
Featuring pre-segmented organs, this dataset is well-suited for non-expert users, such as junior medical school students or the general public, to gain a better understanding of human anatomy.

\begin{figure}[htb]
	\centering
	\includegraphics[width=1.0\linewidth]{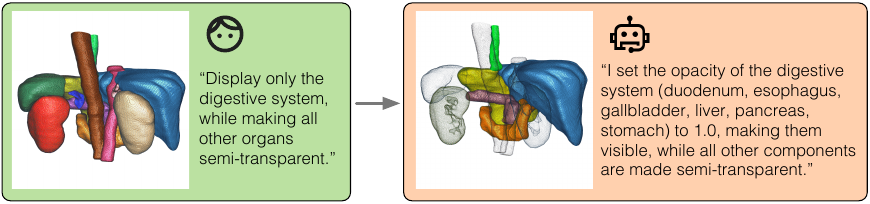}
	\vspace{-0.25in}
	\caption{Case study with the FLARE human organ dataset demonstrating semantic filtering via NLI.}
	\label{fig:flare}
\end{figure}

NLI4VolVis enhances data accessibility and comprehension by enabling semantic filtering and interactive visualization. 
As shown in Figure~\ref{fig:flare}, users can request NLI4VolVis to display only organs related to the digestive system, supporting both exploratory analysis and presentation goals. 
Using the open-vocabulary query tool, our multi-agent system first identifies relevant anatomical structures—including the liver, pancreas, gallbladder, esophagus, stomach, and duodenum. 
It then retrieves the corresponding segmented components and applies declarative commands to generate the requested view.

{\bf Supernova.}
The supernova dataset is a scientific simulation that captures the dynamic processes of a stellar explosion. 
We segment it using a TF into three key components: the expanding shockwave, turbulent plasma, and dense core ejecta.

\begin{figure}[htb]
	\centering
	\includegraphics[width=1.0\linewidth]{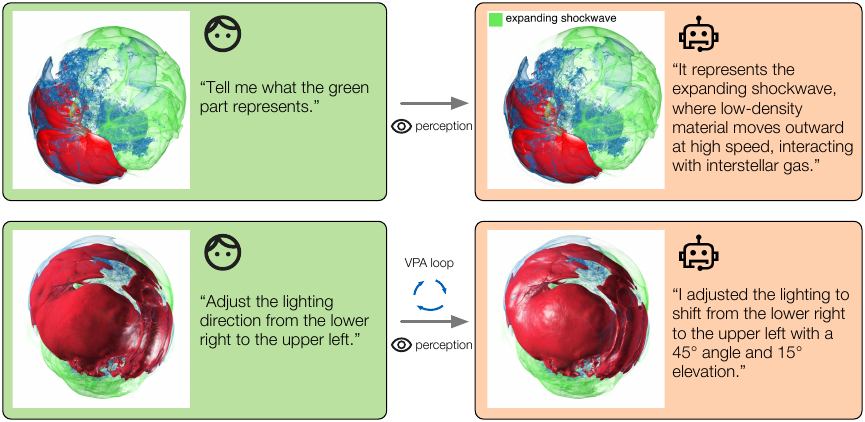}
	\vspace{-0.25in}
	\caption{Case study with the supernova dataset demonstrating the perception capabilities of NLI4VolVis.}
	\label{fig:supernova}
\end{figure}

NLI4VolVis leverages its multi-modal capabilities and the extensive internal knowledge of LLMs to provide users with an interactive and intuitive understanding of complex scientific phenomena. 
As shown in Figure~\ref{fig:supernova}, users request to explain the green region in the visualization. 
Our system first identifies this region as the expanding shockwave using its perception capabilities. 
It then adds a legend to the requested region and generates a contextual explanation of the expanding shockwave, enhancing user comprehension of the dataset.
Users further request an adjustment to the lighting direction. 
NLI4VolVis successfully shifts the lighting from the lower right to the upper left by iteratively applying its perception capability through the VPA loop.

\vspace{-.1in} 

{\bf Backpack.}
The backpack dataset is a CT scan of a backpack containing various everyday objects, such as water bottles and ropes. 
Visualizing such datasets has practical applications, particularly in security screening at airports and train stations.

\vspace{-.1in} 
\begin{figure}[htb]
	\centering
	\includegraphics[width=1.0\linewidth]{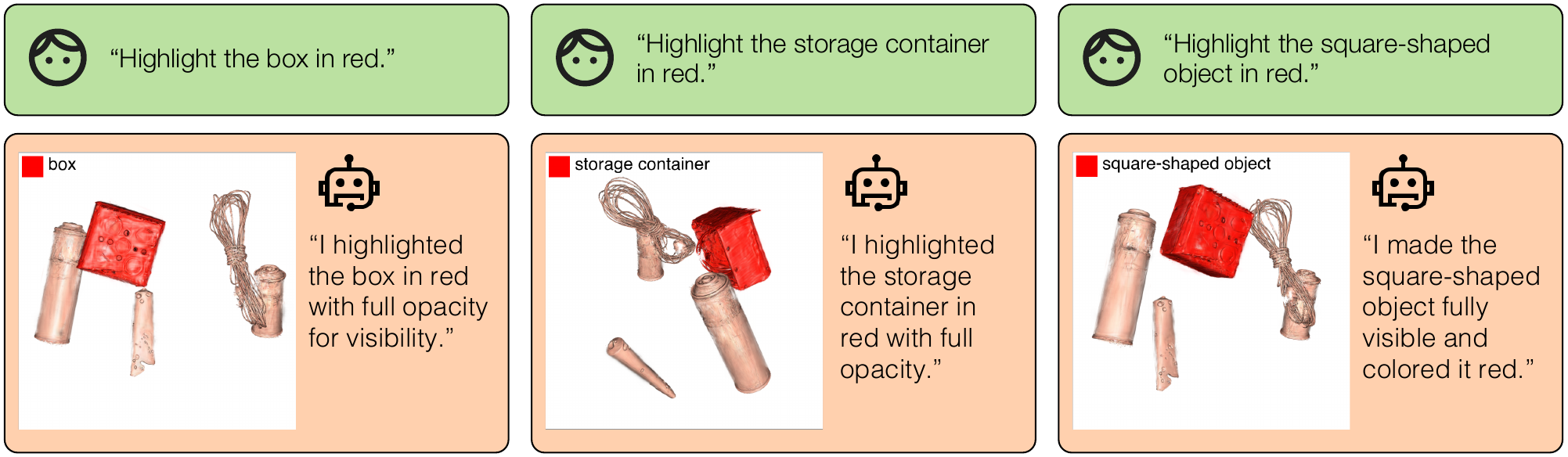}
	\vspace{-0.25in}
	\caption{Case study with the backpack dataset demonstrating how open-vocabulary querying handles vague natural language references.}
	\label{fig:backpack}
\end{figure}

To process the dataset, we first apply a TF to isolate objects within the backpack before segmenting the entire scene into five semantic components. 
As shown in Figure~\ref{fig:backpack}, users interact with the system by referring to the same object using different descriptions, such as ``{\em the box},'' ``{\em the storage container},'' or ``{\em the square-shaped object}.'' 
Despite these variations, our multi-agent system accurately interprets user intent and employs the open-vocabulary query tool to correctly identify the referenced object, demonstrating its robustness in handling vague or context-dependent descriptions.

\begin{figure}[htb]
%\vspace{-.1in} 
	\centering
	\includegraphics[width=1.0\linewidth]{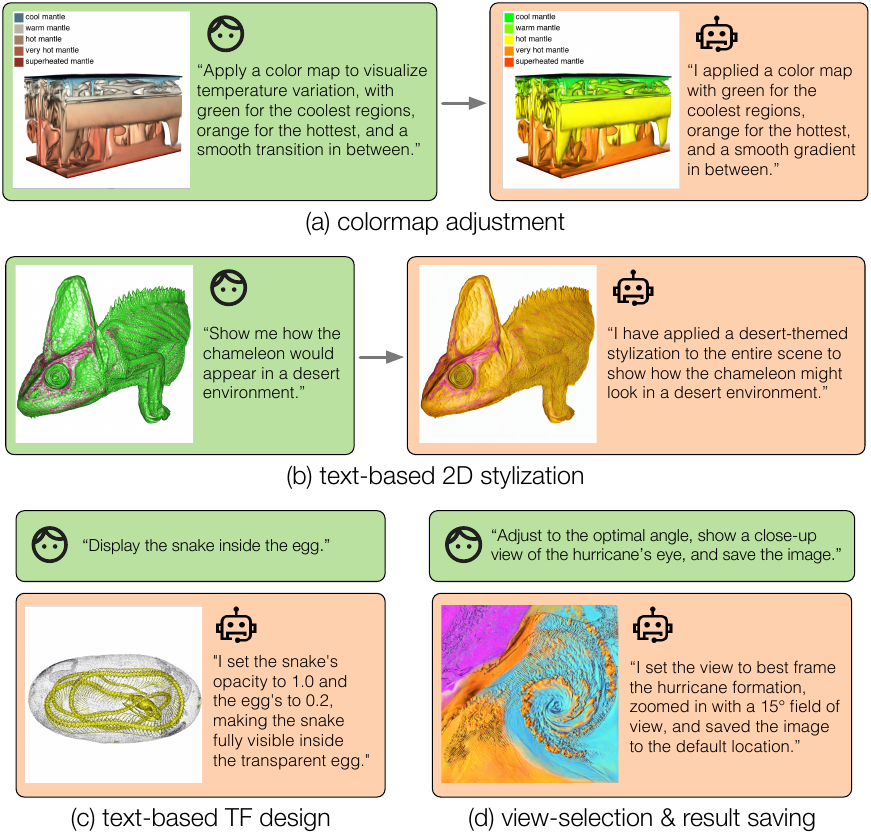}
	\vspace{-0.25in}
	\caption{Representative user interactions across different datasets.}
	\label{fig:others}
\end{figure}

{\bf Other case studies.}
We evaluate NLI4VolVis across all eight datasets listed in Table~\ref{tab:dataset}, demonstrating its adaptability to diverse data types. 
Figure~\ref{fig:others} presents representative user interactions, highlighting various system functionalities applied to different datasets. 
These interactions include colormap adjustments for visualizing temperature variations, text-based 2D stylization for adapting to environmental contexts, text-driven TF design, view selection, and result saving. 
Each case illustrates how users leverage natural language queries to refine visualizations and extract meaningful insights.

\vspace{-.1in} 

\vspace{-0.05in}
\subsection{User Study}
\label{subsec:us}

Our user study evaluates how effectively NLI4VolVis responds to user queries, the effort required for non-experts to familiarize themselves with the system, its impact on users' understanding of volumetric datasets, and how well users can communicate with the system to achieve their visualization goals.

{\bf Participants and experiment setup.}
Eight participants were recruited for the study following the University's IRB protocol. 
The group included three individuals with master's degrees in mathematics or computer engineering and five with bachelor's degrees. 
Among them, six had experience in computer science-related fields, including machine knitting, human-computer interaction, LLM systems, medical image processing, and graph learning. 
The remaining two specialized in biomolecular engineering and chemistry. 
%Three participants had prior knowledge of the datasets used, while six had basic experience with visualization tools. 
\hot{Three participants reported having prior domain knowledge of the datasets due to their academic background; however, none had previously explored these specific datasets using visualization tools before the study. 
The remaining five participants had no prior knowledge of the datasets.}
However, only two were highly familiar with natural language-driven visualization systems. 
Each participant interacted with the system on a workstation equipped with an NVIDIA RTX 4090 GPU and a 32-inch display with a resolution of 2560$\times$1440.

{\bf Procedure and tasks.}
The study followed a four-phase workflow: {\em introduction}, {\em exploration}, {\em task execution}, and {\em post-questionnaire}.
Participants were first introduced to NLI4VolVis through a ten-minute tutorial and a demonstration video using the backpack dataset to illustrate the system's core functionalities. 
They then explored the interface freely with the backpack and chameleon datasets, with the opportunity to ask questions. 
Exploration continued until they felt confident in using the system.

In the task execution phase, participants worked with three datasets: carp (biological dataset), mantle temperature (simulation dataset), and FLARE human organ (medical dataset). 
The study included 23 tasks covering various visualization and interaction challenges:
guided dataset exploration (3 tasks), 
answering domain-specific knowledge questions (9 tasks), 
modifying visualization properties (e.g., color, opacity, lighting) (4 tasks), 
optimizing views (1 task), 
applying 2D stylization (1 task), and 
semantic filtering (e.g., identifying the largest organ) (5 tasks). 
Finally, to evaluate the system's natural language processing capabilities, participants who initially completed tasks using only the {\em graphical user interface} (GUI) were asked to replicate those tasks exclusively through NLI.

%Likert-1932
The study concluded with a post-questionnaire, where participants rated usability, efficiency, and interaction quality using a five-point Likert scale. 
Open-ended feedback was collected to identify system strengths, challenges, and areas for improvement. 
All eight participants completed all four phases, and each was compensated \$30. 
Their interactions were recorded for subsequent analysis, with performance measured based on task completion time, the number of queries required, and response accuracy. 
Additionally, we recorded the time each participant spent exploring the system and the total time taken to complete all tasks.

\vspace{-.1in} 
\begin{table}[htb]
	\caption{User task performance across seven metrics.}
	\vspace{-0.1in}
	\centering
	%{\scriptsize
		\resizebox{\linewidth}{!}{
			\begin{tabular}{r|cccccccc|c}
				metric  &P1  & P2  & P3 & P4 & P5 & P6 & P7 & P8 & mean  \\ \hline
				exploration time (min) & 8 & 24 & 5 & 8 & 4 & 5 & 16 & 9 & 9.9 \\
				task completion time (min) & 34 & 32 & 42 & 32 & 34 & 32 & 40 & 35 & 35.1 \\
				NL interactions & 35 & 26 & 27 & 16 & 20 & 20 & 28 & 23 & 24.4 \\
				effective NL interactions & 28 & 21 & 21 & 15 & 18 & 19 & 24 & 19 & 20.6 \\
				non-NL interactions & 8 & 0 & 1 & 9 & 4 & 2 & 2 & 1 & 3.4 \\
				NL usage rate (\%) & 81.4 & 100.0 & 96.4 & 64.0 & 83.3 & 90.9 & 93.3 & 95.8 & 88.1 \\
				knowledge QA accuracy & 8/9 & 9/9 & 9/9 & 9/9 & 9/9 & 9/9 & 8/9 & 9/9 & 8.75/9 \\
			\end{tabular}
		}
	%}
	\label{tab:task_performance}
	\end{table}

{\bf Task performance.}
Table~\ref{tab:task_performance} summarizes user task performance from the study, with P1 to P8 representing the eight participants. 
The reported metrics include: 
``exploration time''—the duration each participant spent familiarizing themselves with the system before starting the tasks, 
``task completion time''—the total time taken to complete all assigned tasks, 
``NL interactions''—the number of interactions using natural language,
``non-NL interactions''—the number of interactions using other methods, such as the GUI,
``effective NL interactions''—the number of natural language queries that successfully contributed to task completion,
``NL usage rate''—the proportion of interactions conducted via natural language, and
``knowledge QA accuracy''—the number of correctly answered domain-specific questions (out of nine).

All participants completed the tasks, although their interaction patterns and completion times varied. 
Exploration time ranged from 4 to 24 minutes, averaging 9.9 minutes. 
Some participants, such as P2 and P7, spent more time exploring the system. 
However, exploration time did not correlate strongly with task completion time, which averaged 35.1 minutes, with individual times ranging from 32 to 42 minutes.

\begin{figure}[htb]
\vspace{-.1in} 
	\centering
	\includegraphics[width=1.0\linewidth]{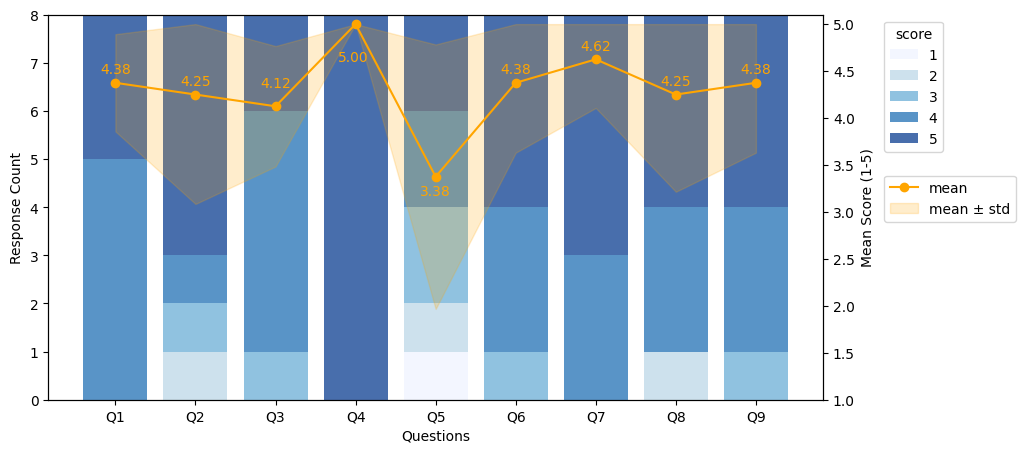}
	\vspace{-0.25in}
	\caption{Survey responses to nine post-questionnaire questions, along with mean scores (on a 1–5 scale) and standard deviations.}
	\label{fig:survey}
\end{figure}

Participants demonstrated differing preferences between NLIs and traditional GUI controls. 
For example, P4, who was highly familiar with interactive visualization systems and had prior experience in computer graphics, relied on NL for only 64\% of interactions, which was notably lower than the average NL usage rate of 88.1\%. 
In contrast, two participants with no computer science background, P2 and P8, used NL almost exclusively, with usage rates of 100\% and 95.8\%, respectively.

All participants answered domain-specific knowledge questions well, achieving an average accuracy of 8.75/9. 
The number of interactions required to complete the tasks varied across participants. 
On average, participants performed 24.4 NL interactions, with 20.6 being effective, and made 3.4 non-NL interactions while completing 23 tasks. 
Since 14 tasks required at least one interaction (excluding knowledge QA tasks), some participants made additional interactions beyond the minimum required. 
Analyzing the recorded interactions revealed that this behavior was driven by two main factors: some participants lacked patience and initiated new interactions before the system responded, while others deliberately explored the system's capabilities.

{\bf Survey responses.}
We assessed the accessibility and usability of NLI4VolVis through a post-questionnaire. 
After completing the tasks, each participant completed a survey that included their background information, ratings of the system on a 1–5 scale across various aspects, a description of one highlight of the system, notes on challenges encountered during the study, and suggestions for improvements. 
The survey results are presented in Figure~\ref{fig:survey}.

Overall, participants gave positive feedback, with an average score of 4.38/5 for Q1 (``{\em The system allows me to efficiently interact with and manipulate volumetric datasets}''). 
Three questions focused on accessibility: Q2 (``{\em I can easily complete visualization tasks without assistance},'' 4.25/5), Q4 (``{\em The system is easy to learn, even for first-time users},'' 5.0/5), and Q8 (``{\em The system's interface is intuitive and user-friendly},'' 4.38/5). 
Most participants found NLI4VolVis highly accessible, especially compared to traditional VolVis tools like ParaView. 
Four participants (P1, P3, P4, and P5) mentioned the system's ease of learning as its most valuable feature.

The survey also assessed usability, with participants rating all aspects 4.1 or higher, except for Q5 (``{\em I find the system's voice input and output features useful for interaction},'' 3.38/5). 
While four participants rated the voice features 4 or 5, two rated them 1 or 2. 
Observations from the user study revealed that voice input usage varied significantly: P1 and P8 primarily used voice commands, while others preferred keyboard input. 
Additionally, some participants found the voice output distracting and chose to mute the system. 
This suggests that while voice interaction is convenient for some users, it may not be universally preferred.

{\bf User behaviors.}
Survey responses indicate that NLI significantly simplifies complex visualization tasks, making them more intuitive. 
One participant noted, ``{\em Most existing visualization tools are difficult for beginners, but this system transforms complex adjustments into natural language, making it very easy to use.}'' 
This was particularly evident in lighting manipulation, which traditionally requires adjusting Blinn-Phong parameters—selecting between headlight or orbital lighting, modifying azimuthal and polar angles, and fine-tuning ambient, diffuse, specular, and shininess coefficients. 
These adjustments require technical knowledge, as increasing the first three parameters strengthens lighting while increasing the last one reduces brightness. 
Before the user study, none of the participants fully understood these settings. However, all were able to achieve their desired lighting effects simply by stating commands such as ``{\em make the scene brighter}'' or ``{\em change the lighting direction to the upper left}.''

When asked about the most useful aspects of the system, participants frequently highlighted the intuitive natural language interface, as well as features such as color and lighting adjustments, the guided tour, and scene segmentation. 
Some also appreciated the convenience of voice input, which reduced the need for manual adjustments.

Despite its strengths, latency emerged as the most frequently mentioned challenge, with several participants reporting slow response times and uncertainty about whether the system was still processing their input. 
Others noted occasional misinterpretation of commands, interface aesthetics, and a need for additional documentation for first-time users. 
One found that the legend was sometimes inaccurate, while another emphasized the need for clearer explanations of traditional rendering techniques for users unfamiliar with computer graphics.

For improvements, participants suggested refining zoom and rotation functions, adding a ``Reset All'' button, and enhancing real-time system feedback. 
Several users requested status indicators to confirm when the system was actively processing their commands. 
One specific suggestion was to display system messages such as ``Processing...'' in the terminal window during execution. 
Participants also suggested adding hover-based tooltips for quick access to information and enhancing visual cues, such as highlighting selected objects and toolbars.
These insights provide valuable feedback on the strengths and areas for improvement in NLI4VolVis, guiding future enhancements to improve efficiency, usability, and system responsiveness. 

{\bf System improvement.} 
In response to participant feedback, we have made several enhancements to NLI4VolVis. 
To address latency issues, we optimized the system by implementing a deque-based memory management strategy for the LLM agents, enabling customization of memory length to balance performance and responsiveness. 
Nevertheless, advanced memory pruning techniques, such as summarization-based approaches~\cite{Maharana-ACL24, Xu-ACL22} or retrieval-based hierarchical methods~\cite{Zhang-EMNLP22}, could further improve efficiency and long-term coherence, which we plan to explore in future work. 
Additionally, we have introduced a ``Reset All'' button for easier scene reinitialization and added real-time processing indicators to inform users when the system is handling commands.

\vspace{-0.05in}
\section{Limitations and Future Work}

{\bf Extension of multi-agent system.} 
NLI4VolVis is designed specifically for {\em cooperative} interactions among agents without incorporating {\em competitive} dynamics. 
Additionally, our collaboration strategy follows a {\em rule-based} approach, where predefined rules dictate agent communication. 
While this ensures coordinated actions, it limits adaptability in uncertain scenarios. 
We plan to explore {\em role-based} and {\em model-based} collaboration strategies to enhance flexibility.
Regarding agent communication, we employ a {\em hierarchical} structure, which is well-suited for question-answering and reasoning tasks. 
However, this design introduces challenges such as increased latency and potential system failure if a higher-level node becomes unresponsive. 
To mitigate these issues, future research could explore {\em centralized} or {\em decentralized} communication models to improve system robustness and reduce latency.

\hot{
{\bf Lack of validation for function-calling tools.} 
NLI4VolVis exhibits strong capabilities in semantic understanding and manipulation of 3D scenes through external function-calling tools. 
However, we observe that the system tends to be overconfident in these tools' outputs and lacks mechanisms for validating their correctness. 
For instance, when a user submits a query unrelated to the current scene, the open-vocabulary querying module still computes the CLIP similarity between the query and all segmented components and returns the top match to the LLM agents. 
For {\em visionless} LLMs (e.g., DeepSeek-V3), there is no means to assess whether the returned result is contextually valid, often leading to inappropriate or misleading visualizations. 
In contrast, {\em vision-enabled} models (e.g., GPT-4o) can detect such mismatches and issue warnings or clarification prompts. 
In future work, incorporating automatic validation techniques within the LLM agent framework will further enhance system robustness and reliability.
}

{\bf Creation of benchmark for VolVis.}
While our user study incorporated carefully designed VolVis tasks, the field still lacks a standardized, large-scale benchmark for quantitatively evaluating LLM agents in SciVis, particularly in volumetric data analysis. 
In contrast, information visualization benchmarks~\cite{Chen-TVCG25, Luo-arXiv25} have begun integrating natural language queries with ground-truth visualizations and automated evaluations, some even introducing ambiguity and multiple valid outputs to assess LLM reasoning under uncertainty.
In future work, we aim to develop a VolVis-specific benchmark featuring diverse datasets, natural language queries, reference visualizations, and automated evaluation of visualization accuracy and interaction quality. 
Such a benchmark would enhance reproducibility and enable more rigorous assessments of NLI-driven VolVis systems like NLI4VolVis.
\hot{In addition, extending the application of NLI4VolVis to a wider range of real-world volumetric datasets beyond our current case studies presents an exciting avenue.}

%% file: conclusion.tex
\vspace{-0.05in}
\section{Conclusions}

We have introduced NLI4VolVis, an NLI framework for VolVis. 
NLI4VolVis integrates multi-view segmentation, vision-language models, and editable 3DGS to extract, interpret, and reconstruct semantic components in volumetric scenes. 
It enables users to explore, query, and manipulate these scenes through an LLM-based multi-agent system with robust function-calling capabilities.
We evaluated NLI4VolVis through case studies across eight datasets, complemented by a formal user study. 
The results show that participants quickly learned the system, engaged naturally through language, and accurately completed assigned tasks. 
By making complex VolVis tasks more intuitive and enhancing accessibility to volumetric data exploration, NLI4VolVis has the potential to democratize VolVis, making it more approachable for both experts and non-experts alike.

%% file: appendix.tex
%\newpage
%\clearpage
\section*{Appendix}

\setcounter{section}{0}
\setcounter{figure}{0}
\setcounter{table}{0}
%\setcounter{page}{1}

%\vspace{-0.05in}
\section{Comparison of Segmentation Methods} 
\label{appendix:segmentation}

\hot{To evaluate the impact of 3D multi-view semantic segmentation methods, we compared LangSplat~\cite{Qin-CVPR24}, SAGD~\cite{Hu-arXiv25}, and SAM 2~\cite{Ravi-arXiv24} using the backpack dataset. 
As illustrated in Figure~\ref{fig:ablation_seg}, each method generates semantic components used to train independent editable 3DGS models, which are subsequently composed into a complete volumetric scene.
LangSplat preserves lighting and surface details effectively but suffers from lower segmentation accuracy, occasionally merging distinct components. 
SAGD achieves the most accurate semantic separation, enabling fine-grained, object-level editing; however, it introduces surface blurring due to its dense Gaussian decomposition. 
SAM 2 provides a middle ground between segmentation accuracy and visual fidelity, offering moderate semantic precision and better temporal consistency than LangSplat, though its reconstructions tend to appear flatter with diminished lighting and texture quality.
Given these trade-offs, we selected SAGD for the backpack dataset to take advantage of its precise object segmentation.

\vspace{-0.1in}
\begin{figure}[htb]
	\centering
	\includegraphics[width=1.0\linewidth]{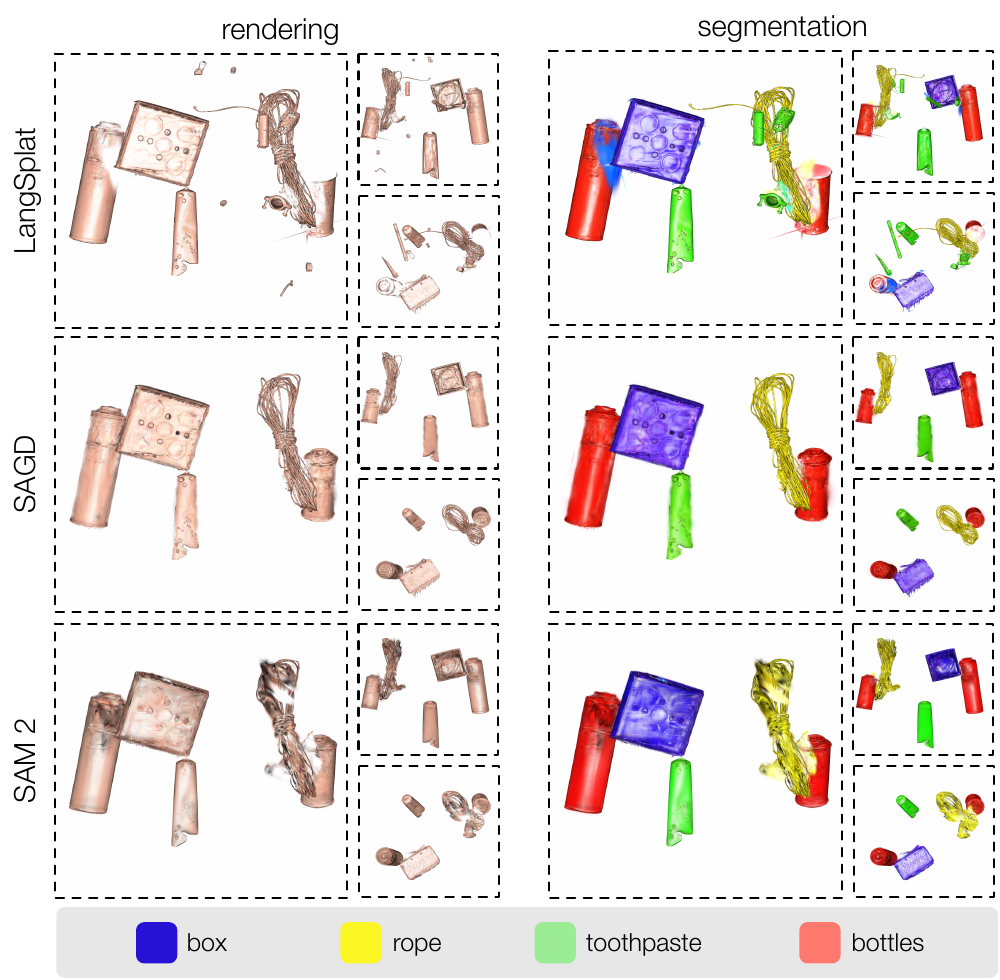}
	\vspace{-0.25in}
	\caption{Comparison of LangSplat, SAGD, and SAM 2 on the backpack dataset. Each row shows the renderings of the reconstructed scene (left) alongside the corresponding semantic segmentation (right).}
	\label{fig:ablation_seg}
\end{figure}

We offer these guidelines for choosing a segmentation method:
\begin{myitemize}
\vspace{-0.05in}
\item \textbf{LangSplat} is best suited for applications that prioritize high visual fidelity and surface detail, where perfect semantic segmentation is not critical.
\item \textbf{SAGD} is ideal for scenarios requiring accurate object-level segmentation to enable precise querying and editing, especially in scenes with well-separated components.
\item \textbf{SAM 2} is a balanced choice when both segmentation accuracy and visual quality are important, particularly for complex scenes where exact surface reconstruction is less essential.
\vspace{-0.05in}
\end{myitemize}
Since no single method is universally optimal, users should select the approach that best aligns with their data characteristics and visualization goals. The modular architecture of NLI4VolVis supports seamless switching between segmentation strategies without requiring changes to the overall pipeline.
}

\vspace{-0.05in}
\section{Comparison of LLMs}
\label{appendix:LLM}

\begin{figure}[htb]
	\centering
	\includegraphics[width=1.0\linewidth]{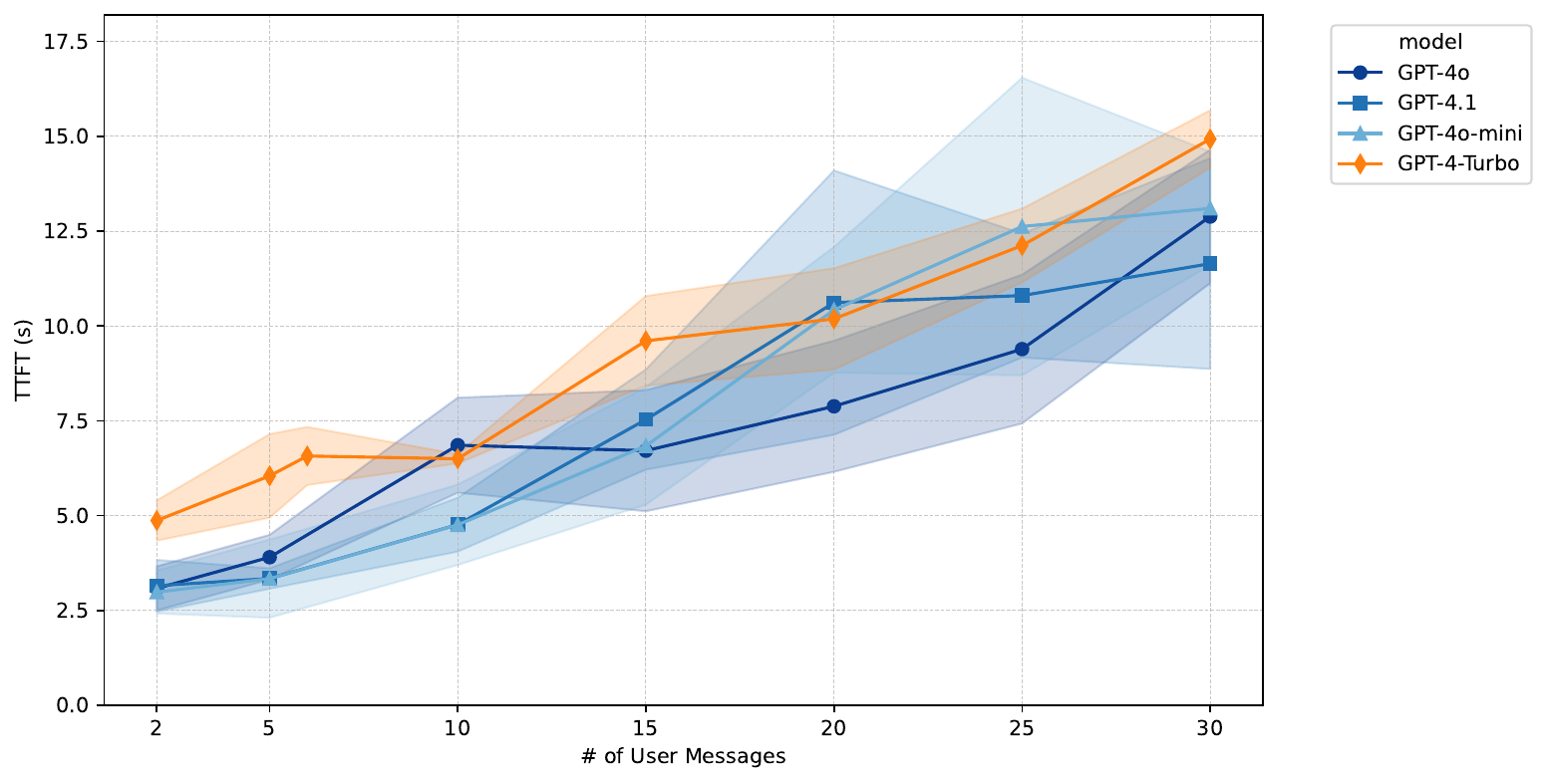}
	\vspace{-0.25in}
	\caption{\hot{Average TTFT for NLI4VolVis plotted against the number of user messages across different GPT models. Shaded regions represent standard deviations.}}
	\label{fig:LLM_TTFT}
\end{figure}

\hot{
To accommodate diverse user preferences and performance requirements, NLI4VolVis supports integration with multiple multimodal LLMs via API. 
Latency among these models varies significantly depending on the provider. 
To quantitatively assess system responsiveness, we measured the {\em time to first token} (TTFT) for four GPT models, varying the number of user messages sent to NLI4VolVis. 
All tests were conducted using the OpenAI API under identical hardware and network conditions. 
For each model and message count combination, we repeated the experiment five times to calculate the mean and variance.

As shown in Figure~\ref{fig:LLM_TTFT} and Table~\ref{tab:TTFT_performance}, TTFT increases with the number of messages, and the variance in latency also grows, leading to less consistent response times. 
In our user study, most participants completed their tasks with fewer than ten messages, keeping per-response latency under 7.5 seconds in most cases. 
Based on these findings, we set the maximum message count in NLI4VolVis to ten, balancing latency and memory efficiency. 
Future enhancements could incorporate memory optimization techniques such as summarization-based methods~\cite{Maharana-ACL24, Xu-ACL22} or retrieval-based hierarchical strategies~\cite{Zhang-EMNLP22} to further improve performance.

Among the GPT models, GPT-4o offers the best overall trade-off between latency, output quality, and cost. 
For other model families (e.g., LLaMA, DeepSeek), we either lacked sufficient local deployment resources or found the available APIs failed to meet low-latency requirements; therefore, TTFT results for these models are not reported.

During the user study, participants freely experimented with GPT-4o, DeepSeek-V3, and LLaMA3.2-90B-Vision. 
GPT-4o consistently delivered superior performance in instruction following and language grounding. 
DeepSeek-V3 exhibited high latency—often exceeding 15 seconds for the first token—even with fewer than five messages, largely due to API limitations. 
LLaMA3.2-90B-Vision struggled with intent interpretation for complex queries, and local deployment was not feasible; its available APIs also suffered from high latency. 
Nonetheless, NLI4VolVis maintains flexibility: users can easily switch LLMs via API configurations to balance output quality and system responsiveness.
}

\vspace{-0.1in}

\begin{table}[htb]
	\caption{\hot{Average TTFT (in seconds) across varying numbers of user messages for each GPT model.}}
	\label{tab:TTFT_performance}
	\vspace{-0.1in}
	\centering
	\resizebox{\linewidth}{!}{
		\begin{tabular}{c|ccccccc}
			%\toprule
			& \multicolumn{7}{c}{\# user messages} \\ 
			%\cmidrule(lr){2-8}
			model & 2 & 5 & 10 & 15 & 20 & 25 & 30 \\ \hline
			%\midrule
			GPT-4o & 3.08 ± 0.58 & 3.90 ± 0.59 & 6.86 ± 1.25 & 6.71 ± 1.60 & 7.88 ± 1.73 & 9.39 ± 1.96 & 12.89 ± 1.76 \\
			GPT-4.1 & 3.15 ± 0.68 & 3.34 ± 0.27 & 4.76 ± 0.71 & 7.53 ± 1.32 & 10.62 ± 3.49 & 10.80 ± 1.64 & 11.64 ± 2.78 \\
			GPT-4o-mini & 2.98 ± 0.57 & 3.34 ± 1.03 & 4.76 ± 1.06 & 6.84 ± 1.55 & 10.42 ± 1.66 & 12.62 ± 3.92 & 13.10 ± 1.50 \\
			GPT-4-Turbo & 4.87 ± 0.53 & 6.05 ± 1.10 & 6.50 ± 0.12 & 9.61 ± 1.18 & 10.19 ± 1.34 & 12.12 ± 0.97 & 14.93 ± 0.76 \\
			%\bottomrule
		\end{tabular}
	}
\end{table}

\vspace{-0.1in}
\begin{figure*}[htb]
	\centering
	\includegraphics[width=1.0\linewidth]{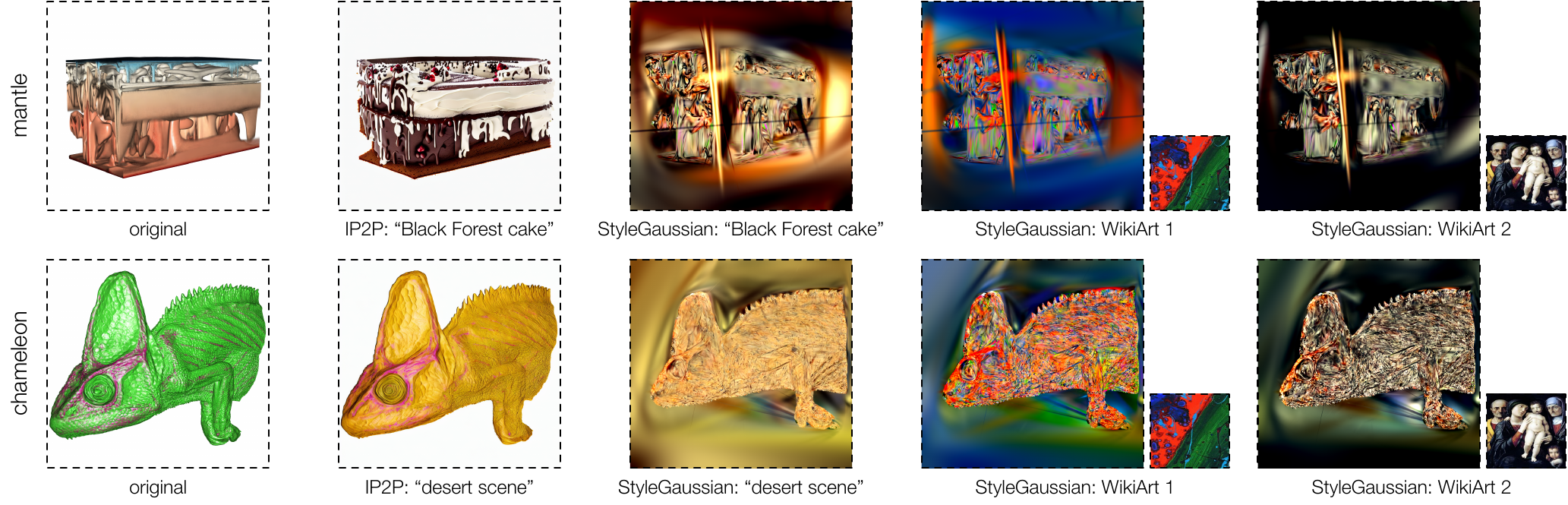}
	\vspace{-0.25in}
	\caption{\hot{2D stylization (IP2P) and 3D stylization (StyleGaussian) results on the mantle and chameleon datasets. The second and third columns show stylizations generated from text prompts, while the last two columns display StyleGaussian's style transfer results using reference images from the WikiArt dataset, shown in the lower-right corners.}}
	\label{fig:stylization}
\end{figure*}

\vspace{-0.05in}
\section{Comparison of 2D and 3D Stylization Methods}
\label{appendix:stylization}

\hot{We compared 2D stylization using IP2P~\cite{Brooks-CVPR23} and 3D stylization using StyleGaussian~\cite{Liu-arXiv24} on the mantle and chameleon datasets. 
Because StyleGaussian does not support text-driven prompts, we first applied IP2P to stylize a single view using textual input, then used the resulting image as a style reference to train StyleGaussian.

Training StyleGaussian on WikiArt took over three hours on an NVIDIA RTX 4090 GPU. 
Inference performance in our tests fell below the reported 10 FPS; we note that the original paper used an RTX A5000 GPU, which outperforms typical consumer-grade hardware. 
As shown in Figure~\ref{fig:stylization}, StyleGaussian introduces noticeable artifacts when applied to SciVis data, such as the mantle example. 
Moreover, like most zero-shot radiance field style transfer methods, it only alters color while preserving geometry, limiting its creative flexibility.
In contrast, IP2P supports prompt-driven stylization but lacks multi-view consistency. 
It completes each stylization in about ten seconds.

To mitigate view inconsistencies, our system temporarily freezes the camera view during IP2P stylization and reminds users to unfreeze it afterward if they wish to resume interactive exploration.
Considering these tradeoffs—high training cost, limited interactivity, visual artifacts on SciVis data, and the lack of text-driven control—we chose not to incorporate StyleGaussian or similar 3D style transfer methods in NLI4VolVis. 
We plan to investigate more advanced and consistent stylization solutions in future work.}

\vspace{-0.05in}
\section{Top-k Frames for View Selection}

\hot{We conducted an ablation study to assess the impact of the parameter \(k\) in our entropy-based view selection strategy. 
In this approach, the top-\(k\) frames with the highest entropy values are selected to generate CLIP embeddings for each segmented semantic component, enabling the 3D open-vocabulary understanding described in Section 3.2 of the paper.

The study was performed on the backpack and carp datasets, which contain five and seven semantic components, respectively. 
For each component, we calculated the similarity between its aggregated top-\(k\) CLIP embedding and the embedding of its ground-truth label. 
We report the mean similarity and variance across all components for each dataset.
As shown in Table~\ref{tab:topk_clip_similarity}, the system exhibits low sensitivity to the choice of \(k\) when \(k \geq 3\). 
However, using only the top-1 frame (\(k{=}1\)) slightly decreases robustness and semantic accuracy, likely due to limited visual context. Conversely, increasing \(k\) leads to longer preprocessing times without noticeable improvements in embedding quality. 
Based on these findings, we recommend setting \(k\) between 5 and 10 to strike a balance between semantic performance and computational efficiency.}

\begin{table}[htb]
	\caption{\hot{Similarity between CLIP embedding and ground-truth label across different top-\(k\) frames for entropy-based view selection.}}
	\label{tab:topk_clip_similarity}
	\vspace{-0.1in}
	\centering
	\resizebox{1.0\columnwidth}{!}{
		\begin{tabular}{c|ccccc}
			%\toprule
			& \multicolumn{5}{c}{top-\(k\) frames} \\ 
			%\cmidrule(lr){2-6}
			dataset & \(k{=}1\) & \(k{=}3\) & \(k{=}5\) & \(k{=}10\) & \(k{=}15\) \\ \hline
			%\midrule
			backpack & 0.748 ± 0.00031 & 0.763 ± 0.00036 & 0.766 ± 0.00034 & 0.764 ± 0.00037 & 0.763 ± 0.00036 \\
			carp     & 0.737 ± 0.00031 & 0.739 ± 0.00026 & 0.741 ± 0.00026 & 0.740 ± 0.00026 & 0.742 ± 0.00028 \\
			%\bottomrule
		\end{tabular}
	}
\end{table}